\documentclass[useAMS,usenatbib,usegraphicx]{mn2e}

\usepackage{multirow}

\title[Compact binary Galactic dynamics]{The long and the short of it: modelling double neutron star and collapsar Galactic dynamics}
\author[P.D. Kiel, J.R. Hurley and M. Bailes]{Paul D. Kiel$^{1}$\thanks{E-mail:
pkiel@astro.swin.edu.au (PDK)}, Jarrod R. Hurley$^{1}$ and Matthew Bailes$^{1}$\\
$^{1}$Centre for Astrophysics and Supercomputing, Swinburne University of Technology, Hawthorn, Victoria, 3122, Australia}

\begin{document}


\pagerange{\pageref{firstpage}--\pageref{lastpage}} \pubyear{2006}

\maketitle

\label{firstpage}


\begin{abstract}

Understanding the nature of galactic populations of double compact binaries 
(where both stars are a neutron star or black hole) has been a topic of 
interest for many years, particularly the coalescence rate of these binaries.  
The only observed systems thus far are double neutron 
star systems containing one or more radio pulsars.
However, theorists have postulated that short
duration gamma-ray bursts may be evidence of coalescing double neutron 
star or neutron star-black hole binaries, while long duration
gamma-ray bursts are possibly formed by tidally enhanced rapidly rotating 
massive stars that collapse to form black holes (collapsars).
The work presented here examines populations of double compact binary
systems and tidally enhanced collapsars.
We make use of \textsc{binpop} and \textsc{binkin},
two components of a recently developed population synthesis package.
Results focus on correlations of both binary and spatial 
evolutionary population characteristics.
Pulsar and long duration gamma-ray burst observations are used 
in concert with our models to draw the conclusions that: 
double neutron star binaries can merge rapidly on timescales 
of a few million years (much less than that found for 
the observed double neutron star population), 
common envelope evolution within these 
models is a very important phase in double neutron star formation, 
and observations of long gamma-ray burst projected distances are more
centrally concentrated than our simulated coalescing double neutron star 
and collapsar Galactic populations.
Better agreement is found with dwarf galaxy models although
the outcome is strongly linked to the assumed birth radial distribution.
The birth rate of the double neutron star population in our models range from 
$4-160~$Myr$^{-1}$ and the merger rate ranges from $3-150~$Myr$^{-1}$.
The upper and lower limits of the rates results from including electron capture 
supernova kicks to neutron stars and decreasing the common envelope efficiency
respectively. 
Our double black hole merger rates suggest that black holes should receive 
an asymmetric kick at birth.

\end{abstract}


\section{Introduction} 
\label{s:intro}

Pulsars, magnetic oblate spheriods of nuclear densities 
$20-30$ kilometers in diameter, have been found 
rotating at speeds of up to almost one thousand 
times a second (Hessels et al. 2006; see also Galloway 2008) 
in some of the most exotic settings in the known Universe.
For example, some pulsars are found within 
X-ray binaries (Liu, van Paradijs \& van den Heuvel 2007)
and others with compact object companions in close binaries 
(Hulse \& Taylor 1975) thought to be emitting gravitational radiation 
(see Landau \& Lifshitz 1951; 
Paczy\'{n}ski 1967; Clark \& Eardley 1977).
Recently, the number of known binary pulsars has been rapidly 
increasing (Lorimer et al. 2006a; Galloway et al. 2008).
There are now in excess of $100$ 
Galactic disk binary pulsars.
This includes rotation powered pulsars (radio pulsars: ATNF Pulsar 
Catalogue Manchester, Hobbs, Teoh \& Hobbs 
2005\footnote{http://www.atnf.csiro.au/research/pulsar/psrcat/})
and accretion powered pulsars (X-ray binary pulsars: 
Liu, van Paradijs \& van den Heuvel 2007; Galloway 2008; 
Galloway et al. 2008).
More than $20$ of these are accreting from a range of stellar masses and 
companion types (Galloway 2008), 
$9$ are thought to orbit another neutron star 
(van den Heuvel 2007; Stairs 2008), while others ($> 70$) dwell 
in detached systems with 
white dwarfs companions (ATNF Pulsar Catalogue, 
Manchester et al. 2005) and possibly main sequence (MS) 
companions (Champion et al. 2008).
From observations of pulsars in such systems it is possible
to constrain computational modelling of binary evolutionary 
phases that are general to many non-pulsar related systems 
including, but not limited to, tidal evolution, Roche-lobe 
overflow and common envelope (CE) evolution.
It is the most rapidly rotating pulsars that best
constrain uncertainties in the theory related to 
these processes.
Such work has been attempted in Kiel et al. (2008)
and Kiel \& Hurley (2009: KH09), where models
of the complete Galactic pulsar population have been made.
However, we note that these models are yet to include selection effects which are critical when interpreting 
the results of pulsar surveys (Taylor \& Manchester 1977; Oslowski et al. 2009). 

Short gamma-ray bursts provide an alternative method to 
study the physics of compact stars.
Here compact stars are considered to be the most compact of
remnants: neutron stars (NSs) and black holes (BHs).
We define a double compact binary (DCB) to be any combination of these compact 
objects within a binary system (without any limit
on the orbital period) and close DCBs are those
systems with orbital periods of less than a few days.
The DCBs that merge within the assumed $10~$Gyr age of the Galaxy (i.e. during
our simulations) are defined as coalescing DCBs.
It is postulated that gamma-ray emission is produced during
the coalescence of these systems and that radiation of 
gravitational waves occurs during the preceding in-spiral 
phase (Clark \& Eardley 1977).
In particular it is thought that short gamma-ray bursts
are produced by coalescing double neutron star systems 
(Paczynski 1986).
Gamma-ray burst observations are very interesting in themselves, 
however, DCB systems offer other observational features of importance.
Not only can many tests of general relativity be performed 
but they offer insights into a host of observable phenomena 
(e.g. Taylor, Fowler \& McCulloch 1979; Lyne et al. 2004).
The formation of close DCBs requires two stars of 
sufficient mass to interact gravitationally, triggering
mechanisms to decrease the separation between them during
their stellar lifetimes.
What seems to be the most important binary evolutionary 
mechanism in forming close DCBs is the common-envelope 
phase (Paczy\'{n}ski 1976) 
where the evolution following the first supernova (SN) generally 
requires at least one such event. 
The modelling of the CE phase is associated with much uncertainty 
(see, for example, Dewi, Podsiadlowski \& Sena 2006; Belczynski et al. 2007a) 
and  will be discussed further in the following section. 
Interestingly, Voss \& Tauris (2003) found that $1/3$ of BH-NS DCB systems can be 
formed via the direct-SN mechanism (Kalogera 1998) and therefore 
a CE phase is not required in all cases.
Adding further uncertainty to the mix Brown (1995) suggested that unless the initial 
mass ratio is very close to unity
NSs spiraling-in within a CE should always accrete enough matter to collapse
and form BHs.

This work examines population characteristics
of DCBs, including predictions of 
double neutron star (NS-NS) distributions and correlations between 
orbital properties and location.
The results are extended to detail both long and short gamma-ray 
bursts (GRBs) and their progenitors, modelling tidally 
enhanced collapsars and coalescing DCBs.
In particular, projected distances from the host galaxy of model GRBs
are compared to observations.
No detailed conclusions from direct comparison 
to observations are attempted, as in 
Belczynski, Bulik \& Rudak (2002) 
who account for redshift and different galaxy masses.
Nor do we consider the evolution of systems in globular clusters 
(see e.g. Ivanova et al. 2008; Sadowski et al. 2008).
Section~\ref{s:bpbk} briefly outlines the population 
synthesis tool used in this body of work.
The results are spread over Sections~\ref{s:DCBNS-NS}, 
\ref{s:coal} and \ref{s:grb}, which examine the 
bound DCB and NS-NS populations (even if they go on to 
eventually coalesce), coalescing DCB populations
and GRB populations, respectively.

\section{Population synthesis models}
\label{s:bpbk}

The following work utilizes theoretical studies into NS
binary and stellar evolution (Kiel et al. 2008)
and Galactic kinematics (KH09).
The modelling of stellar and binary evolution is performed
by \textsc{binpop}, a Monte-Carlo population synthesis 
scheme that incorporates the 
detailed binary stellar evolution (\textsc{bse}) code, 
developed by Hurley, Tout \& Pols (2002) and updated in 
Kiel et al. (2008).
\textsc{bse} facilitates modelling of the latest 
theoretical stellar and binary evolution 
prescriptions such as: tides, stellar composition, 
magnetic braking, gravitational radiation, 
mass accretion over a range of timescales and 
realistic angular mometum evolution of pulsar systems.
All of which are important evolutionary features 
when modelling populations of DCBs.

An aspect of binary evolution crucial to the formation 
and characteristics of close DCBs  
is common-envelope evolution 
(Paczy\'{n}ski 1976; Webbink 1984; Iben \& Livio 1993). 
The treatment of CE evolution within \textsc{binpop}/\textsc{bse} has 
previously been described in detail 
(Hurley, Tout \& Pols 2002; Kiel \& Hurley 2006). 
However, considering the potential influence of this phase on the outcomes 
of close binary population synthesis we also give an overview here so as to 
assist the reader in the interpretation of our results, noting that a full investigation 
of how uncertainties involved in modeling this phase affect the properties of the 
DCB populations will be left to future work. 
A CE phase is assumed to be initiated if a giant star 
(with a sizeable convective envelope) fills its Roche-lobe and the ensuing 
mass-transfer is calculated to occur on a dynamical timescale. 
The envelope of the giant then rapidly expands and envelops both the 
companion star and the degenerate core of the giant, thus forming a CE. 
As suggested by Paczy\'{n}ski (1976), the companion star and giant core 
will then spiral towards each other while transferring orbital energy to the 
envelope via dynamical friction. 
The eventual outcome of this process is essentially determined by a race 
between the spiral-in and the removal of the envelope owing to the 
energy transfer: 
if the companion and the giant core come into contact first 
then a merger results, otherwise the result is a close binary 
composed of the former core (now a remnant star) and the companion. 
Although some detailed hydrodynamic models of this process have 
been attempted (Bodenheimer \& Taam 1984; Taam \& Sandquist 2000; 
Ricker \& Taam 2008) the treatment within population synthesis 
codes remains simplistic as a detailed theory is lacking. 
In \textsc{binpop}/\textsc{bse} modelling of the CE phase is summed up by the 
equation: 
\begin{equation} 
\frac{M \left( M - M_{\rm c} \right)}{\lambda R} 
= \frac{\alpha_{\rm CE} M_{\rm c} m}{2} \left( \frac{1}{a_{\rm f}} - \frac{1}{a_{\rm i}} \right) \, 
\end{equation} 
which describes the energy balance between the binding energy of the 
envelope and change in orbital energy. 
These are related by an efficiency parameter, $\alpha_{\rm CE}$, 
which encapsulates the uncertainty of the model within what is known
as the $\alpha$-formalism (Nelemans \& Tout 2005).
Also included in the calculation are the mass of the giant star, $M$, 
the mass of the giant core $M_{\rm c}$, the mass of the companion, 
$m$, the gravitational constant, $G$, the structure 
constant of the giant, $\lambda$, the radius of the giant, $R$, 
and the initial and final orbital separation, $a_{\rm i}$ and $a_{\rm f}$. 
As documented 
in Hurley, Tout \& Pols (2002) and Kiel \& Hurley (2006), this is roughly the 
equivalent of using $\alpha^{'}_{\rm CE} = 1$ in the alternative formulation 
given by Iben \& Livio (1993: where we have taken the liberty of using 
$\alpha^{'}_{\rm CE}$ to distinguish between the two methods) 
and thus corresponds to efficient transfer of energy within the CE. 
Kiel \& Hurley (2006) found that using $\alpha_{\rm{CE}} = 3$ aided 
comparison of the relative NS-low-mass X-ray binary and 
BH-low-mass X-ray binary formation rates to observations.
Previous use of $\alpha_{\rm CE} = 1$ produced a dearth 
of BH-low-mass X-ray binaries
compared to NS-low-mass X-ray binaries.

Within the $\alpha$-formalism the greater the value of 
$\alpha_{\rm CE}$ the more efficient the spiral in process 
is at driving away the envelope.
Values greater than unity may suggest an energy source other 
than gravitational potential energy is also important in 
driving off the envelope.
The effect of the assumed value of $\alpha_{\rm CE}$ 
(or $\alpha^{'}_{\rm CE}$) on the
results of binary population synthesis have been documented
previously (e.g. Tutukov \& Yungleson 1996; Belczynski, Bulik \& Rudak 2002;
Hurley, Tout \& Pols 2002; Voss \& Tauris 2003; Kiel \& Hurley 2006).
A large $\alpha_{\rm CE}$ prescription is also equivalent to 
a large value of $\lambda$ -- a diffuse stellar envelope.
In the past there have been three methods of accounting for the
structure of the giant donor star with $\lambda$ (see Voss \& Tauris 2003
for a similar discussion on this point).
One method is to include it with the CE efficiency and to simply vary 
a combined $\alpha_{\rm CE} \lambda$ parameter 
(e.g. Belczynski et al. 2002a,b; 
Nelemans \& Tout 2005; Pfahl, Podsiadlowski \& Rappaport 2005;
Belczynski et al. 2007a).
Another method is to assume a constant value of $\lambda$ 
(with $0.5$ typically used) 
separate to the assumed $\alpha_{\rm CE}$ value (numerically consistent
with the previous method: Portegies Zwart \& Yungleson 1998; 
Hurley, Tout \& Pols 2002; Dewi, Podsiadlowski \& Sena 2006; 
Kiel \& Hurley 2006).
The final method is to include an algorithm which provides
varying values of $\lambda$ depending upon (see Tauris \& Dewi 2001) 
the mass (core and envelope) and stellar evolutionary phase of the donor 
star (e.g. Voss \& Tauris 2003; Podsiadlowski, Rappaport, Han 2003; 
Kiel \& Hurley 2006).
Tauris \& Dewi (2001) show that $\lambda$ values may range between 
$0.02-0.7$ (the average for massive stars being $\lambda = 0.1$: 
Dewi \& Tauris 2001; Dewi, Podsiadlowski \& Sena 2006).
In this body of work we use a variable $\lambda$ that ranges from
$0.01-0.5$ depending upon the donor type and we assume 
$\alpha_{\rm CE}$ to be either $1$ or $3$.

Another feature of the CE phase which has been shown to clearly affect
the resultant populations of double compact binaries is the possibility
that CE initiated by stars on the Hertzsprung Gap (HG) always leads
to coalescence (Belczynski et al. 2007a).
Following this assumption Belczynski et al. (2007a) found that the merger
rate of BH-BH binaries decreased by $\sim 2$ orders of magnitude
compared to models without such a restriction.
This concept is based on the study of Ivanova \& Taam (2003) who, in
agreement with the suggestion of Podsiadlowski, Rappaport \& Han (2003)
found that all their detailed models in which a HG star initiates
CE resulted in coalescence of the two cores.
At this stage 
we do not impose such an outcome explicitly. 
However, the possibility of coalescence is increased by 
the use of a varying $\lambda$ during CE.
Here the value of $\lambda$ may drop to as low as $\sim 0.01$ during the 
HG phase (e.g. Podsiadlowski, Rappaport \& Han 2003) accounting for the 
lack of a convective envelope especially for those stars near the
Hayashi track.

During the rapid and explosive formation of compact objects impulsive asymmetric 
mass-loss can lead to the transmission of momentum to the central compact object.
To model this transfer of momentum we assume an instantaneous SN velocity kick for NSs
in the form of a Maxwellian distribution with a dispersion of $190~{\rm km s}^{-1}$
(see Kiel et al. 2008).
For BH systems there are difficulties in statistically determining a realistic description 
for the SN kick magnitudes of the population. 
The magnitude of stellar BH velocity kicks is believed to vary between
$0$ and $\sim 300~{\rm km s}^{-1}$ (e.g. Jonker \& Nelemans 2004;
Willems et al. 2005).
To account for this uncertainty we produce models where BHs receive no kicks 
or where BH kicks are drawn from a Maxwellian distribution with
a dispersion of $190~{\rm km s}^{-1}$.
We also can include electron capture SN (EC SN; Miyaji et al. 1980) 
events in our models.
The capture of electrons onto primarily magnesium atoms within an 
oxygen-neon-magnesium core depletes the core of electron pressure
resulting in the collapse and supernova explosion.
The details of our EC SN model are found in Kiel et al. (2008) and here we
provide models with and without this assumption.
During EC SNe the SN kick is taken from a Maxwellian distribution with a dispersion assumed to be 
$20~{\rm km s}^{-1}$ (typical values range anywhere from $0-50~{\rm km s}^{-1}$).
The lower dispersion value arises because of the lower energy yield of
EC SNe as compared to typical SNe.

Modelling of stellar and binary kinematic evolution
within a galaxy is completed using \textsc{binkin}.
This code, developed in KH09, solves four
coupled equations of motion to evolve forward in time
the galactic positions of systems of interest.
\textsc{binkin} uses a recipe for the galactic
gravitational potential structure, initial
birth positions, initial system velocities 
and output from \textsc{binpop} to calculate
the kinematic details for a population of stellar
systems within a galaxy.
The particular output from \textsc{binpop} of interest
for \textsc{binkin} is information on SN recoil 
velocities (time, direction and magnitude) and the
birth time of each system within the Galaxy (randomly
selected from a flat distribution between $0-10~$Gyr). 
Previously, both \textsc{binpop} (Kiel et al. 2008) and \textsc{binkin} (KH09)
were used to examine pulsar systems exclusively.
DCB and GRB systems, on the other hand, have not been considered
directly or in such detail by us until now.
We outline the modelling of GRBs systems below and consider their 
population characteristics in Section~\ref{s:grb}.

Gamma-ray bursts are found to occur on two timescales.
There are long GRBs (LGRBs), where the observed bursting event occurs
over timescales of typically $20$s, and short GRBs (SGRBs), with
a median burst duration of $0.3$s (Woosley \& Bloom 2006).
The majority of population synthesis works have examined the
evolution of what is believed to be SGRB progenitors -- coalescing
DCB systems (Portegies Zwart \& Yungelson 1998; 
Bloom, Sigurdsson \& Pols 1999; Hurley, Tout \& Pols 2002; 
Belczynski, Bulik \& Rudak 2002; Voss \& Tauris 2003; 
Belczynski et al. 2008; Sadowski et al. 2008) and similarly 
Paczy\'{n}ski (1990) modelled GRBs taking their locations as those of
an old NS population.
At present it is believed that LGRBs are formed during 
core-collapse type Ibc supernovae (SNe) of massive stars 
whose cores at the time of explosion are rapidly 
rotating (Woosley 1993; Woosley \& 
Bloom 2006; termed as the `collapsar' model by
MacFadyen \& Woosley 1999).
Models suggest that the fall back accretion disks 
of SNe are more energetic and have longer timescales
than models of double compact coalescence
(Woosley 1993; Woosley \& Bloom 2006). 
This leads to a longer GRB event for the former. 
Some alternative models of progenitor SGRBs are outlined
in the review of Nakar (2007; and references within).
These models include accretion induced collapse of NSs into
BHs, type Ia SNe, magnetar giant flares and phase conversion
from NSs to quark or hyper-stars.

Recently, realistic population synthesis 
models of GRBs have examined the formation and 
evolution of LGRBs produced from Population II progenitors 
(see Bogomazov, Lipunov \& Tutukov 2008 and 
Detmers, Langer, Podsiadlowski \& Izzard 2008) and Population III 
progenitors (Belczynski et al. 2007b).
It has been shown that magnetic fields extract
angular momentum from Wolf-Rayet (LGRB progenitor) cores,
slowing the spin of the core below the rate required by GRB theory 
(Petrovic, Langer, Yoon \& Heger 2005; Woosley 1993).
Detailed modelling can produce rapidly rotating 
Wolf-Rayet cores at collapse if the initial stellar metallicity
is reduced below solar quantities (Yoon, Langer \& Norman 2006).
However, in terms of population synthesis both 
Bogomazov, Lipunov \& Tutukov (2008) and 
Detmers et al. (2008) assumed that the formation of LGRBs 
must have occurred within a binary system.
Here a companion star can exert a tidal influence on the 
Wolf-Rayet core, keeping the system tidally locked until the
SN and formation of LGRB and black hole.
Bogomazov, Lipunov \& Tutukov (2008) examined in some detail the
effect modifying the metallicity and wind mass loss had on the
formation of LGRBs, in line with the detailed models of 
Yoon, Langer \& Norman (2006).
Detmers et al. (2008) investigated, via detailed 
stellar models, whether the effect of tides in 
close binary systems at solar metallicities can 
spin up a Wolf-Rayet star sufficiently enough 
to form a collapsar-LGRB.
Although Detmers et al. (2008) produced 
collapsar systems they found it could only arise from 
those systems that had some mass transfer 
and/or merger event during their lifetimes. 
Detmers et al. (2008) then perform a population 
synthesis study to compare their model birthrate
to observations.
They found that the tidally spun-up collapsar model
could only produce a small fraction of the LGRB 
formation rate, however, carbon-oxygen and/or 
helium star mergers with BHs occur more readily and
could possibly form some fraction of GRB systems.

For our LGRB models we follow the method of 
Detmers et al. (2008) which requires a system 
that contains a carbon-oxygen star to be tidally 
influenced by a BH.
The rapidly rotating (tidally locked) carbon-oxygen 
star then goes on to form a BH and in the process a LGRB.
It is not relevant if the resultant BH-BH system becomes
disassociated.
We consider a binary system to be tidally locked when
the ratio of carbon-oxygen star radius to the orbital 
separation is greater than $0.2$ (Portegies Zwart \& 
Verbunt 1996).

The focus of this paper is on DCB and GRB systems produced in 
population synthesis models.
To compare with our previous pulsar population synthesis
work we provide analysis of Model C$^{'''}$ from KH09. 
This model was introduced briefly in KH09 but is used
in detail here.
It involves the evolution of $10^{9}$ binary systems
which is two orders of magnitude greater than the
main suite of models presented in KH09.
Model C$^{'''}$ incorporates the favoured binary 
evolutionary model of Kiel et al. (2008),
Model Fd, and the favoured kinematics of KH09. 
Details for the model can be found
in Table 1 of Kiel et al. (2008) and Table 1 of KH09.
For clarity we show the full set of possible parameters 
in \textsc{binpop} and \textsc{binkin} within 
Table~\ref{t:binpopbinkin} along with their values
for Model C$^{'''}$. 
To address some possible deficiencies of Model C$^{'''}$ we 
also examine the resultant DCB and GRB populations of an additional set of models that include 
different evolutionary assumptions.
The model changes include assuming: 
(i) $\alpha_{\rm CE} = 1$; 
(ii) BHs receive velocity kicks at birth (from the same SN kick distribution as NSs); and, 
(iii) BHs receive kicks \textit{and} NSs may form via EC SN. 
Including these effects allow us to determine the extent by which such assumptions
affect the final model DCB and LGRB populations.
The initial number of binary systems in these additional models is $10^8$,
an order of magnitude less than in Model C$^{'''}$.
As shown in KH09 this decrease in model initial binary number still provides statistical 
significant results for binary evolution and Galactic kinematics, 
while being preferable for computational efficiency. 
In our models we define the Galactic region of interest 
to include all stars with $R \leq 30~$kpc and $|z| \leq 10~$kpc.
We note that when discussing DCB systems the order of 
compact star formation is depicted by the placement
within the system name.
For example the NS is created first in a NS-BH binary
(i.e. before any SN the initially more massive star transfers 
much of its mass to the initially less massive star facilitating 
NS formation prior to BH formation).

\begin{table*}
 \centering
 \begin{minipage}{140mm}
  \caption{
    Model C$^{'''}$ \textsc{binpop} and \textsc{binkin} parameters.
    See also Table 3 of Hurley, Tout \& Pols (2002),
    Table 1 of Kiel \& Hurley (2006), Kiel, Hurley, Murray \& 
    Hayazaki (2007), Table 1 of Kiel et al. (2008) and 
    Table 1 of KH09.
    Note, the references within the table are: 
    Kroupa, Tout \& Gilmore (1993; KTG93),
    Belczynski, Kalogera \& Bulik (2002; BKB02),
    Tauris \& Manchester (1998; TM98), Yusifov \& Kucuk (2004, YK04)
    and Paczy\'{n}ski (1990, Pac90).
    In this work we examine the effect changing some of these parameters have
    on the resultant population, in particular we
    include BH supernova kicks and electron capture
    SNe as well as decreasing the efficiency of the CE phase to $\alpha_{\rm CE} = 1$.
  \label{t:binpopbinkin}}
  \begin{tabular}{clrc}
  \hline \hline
 \textsc{binpop} & & &  \\
 \hline \hline
  \multirow{2}{*}{Parameter} & \multirow{2}{*}{Description} & Value/ & Varied in \\
   & & choice & Kiel et al. (2008)\\
  \hline
  $Z$ & Zero-age main sequence metallicity & $0.02$ & $\times$ \\
  $M_{\rm 1i,range}$ & Primary star birth mass range ($M_{\rm 1i,min} - M_{\rm 1i,max}$)& $5 - 80~M_\odot$ & $\times$ \\
  $\xi(M_1)$ & Birth distribution of primary masses  & KTG93 & $\times$ \\
  $M_{\rm 2i,range}$ & Secondary star birth mass range ($M_{\rm 2i,min} - M_{\rm 2i,max}$)& $0.1 - 80~M_\odot$ & $\times$ \\
  $\phi(M_2)$ & Birth distribution of secondary masses  & $n(q) = 1$ & $\times$ \\
  $P_{\rm orbi,range}$ & Orbital period birth range ($P_{\rm orbi,min} - P_{\rm orbi,max}$)& $1 - 3000~$days & $\times$ \\
  $ \Omega(\log \rm{P}_{\rm orb})$ & Birth distribution of orbital period  & Flat & $\times$ \\
  $e_{\rm i}$ & Initial eccentricity distribution & $0$ & $\times$ \\
  $\alpha_{\rm CE}$ & Common envelope efficiency parameter & $3$ & $\times$ \\
  $\lambda$ & Binding energy factor for common envelope evolution & Variable $\left(0.01 - 0.5 \right)$ & $\times$ \\
  $M_{\rm NS,max}$ & Maximum NS mass & $3~M_\odot$ & $\times$ \\
  $\eta$ & Reimers mass-loss coefficient & $0.5$ & $\times$ \\
  $\mu_{\rm nHe}$ & Helium star wind mass loss efficiency parameter & $0.5$ & $\times$ \\
  $B_{\rm eml}$ & Binary (only) enhanced mass loss parameter & $0.0$ & $\times$ \\
  $\beta_{\rm wind}$ & Wind velocity factor & $1/8$ & $\times$ \\
  $\alpha_{\rm wind}$ & Bondi-Hoyle wind accretion factor & $3/2$ & $\times$ \\
  $\Xi_{\rm wind}$ & Wind accretion efficiency factor & $1.0$ & $\surd$ \\
  $\epsilon_{\rm nova}$ & Fraction of accreted matter retained in nova eruption & $0.0001$ & $\times$ \\
  $\rm{E}_{\rm FAC}$ & Eddington limit factor for mass transfer & $1.0$ & $\times$ \\
  $\rm{T}_{\rm flag}$ & Activates tidal circularisation & `on' & $\times$ \\
  $\rm{BH}_{\rm flag}$ & Allows velocity kick at BH formation & `off' & $\times$ \\
  $\rm{NS}_{\rm flag}$ & Takes NS/BH mass from BKB02 & `on' & $\times$ \\
  $\rm{Be}_{\rm flag}$ & Allows Be star evolution and sets size of Be circumstellar disk & `off' & $\times$ \\
  $\rm{Be}_{\rm method}$ & Sets method used in Be/X-ray evolution: wind ($< 0$) or RLOF ($> 0$) & N/A & $\times$ \\
  $\rm{Be}_{\rm \dot{M}}$ & Mass loss rate from Be star $(10^{-9} - 10^{-12}~M_\odot/\rm{yr})$ & N/A & $\times$ \\
  $\tau_{\rm B}$ & Pulsar magnetic field  decay timescale & $2~$Gyr & $\surd$ \\
  $k$ & Pulsar magnetic field decay parameter during accretion & $3000$ & $\surd$ \\
  $\rm{P}_{\rm flag}$ & Allows propeller evolution & `off' & $\surd$ \\
  $\rm{EC}_{\rm flag}$ & Triggers electron capture supernova evolution & `off' & $\surd$ \\
  $f(\rm{P})$ & Beaming fraction of pulsar according to TM98 & `on' & $\surd$ \\
  $\rm{DL}_{\rm flag}$ & Allows pulsar death line & `on' & $\surd$ \\
  $\rm{SN}_{\rm link}$ & Initial pulsar parameters linked to supernova & `on' & $\surd$ \\
  $P_{\rm 0min}$\footnote{For the SN link models, this value is the average initial spin period, $P_{\rm 0av}$, not the minimum} & Initial minimum pulsar spin period & $0.02~$s & $\surd$ \\
  $P_{\rm 0max}$ & Initial maximum pulsar spin period & $0.16~$s & $\surd$ \\
  $B_{\rm 0min}$ & Initial minimum pulsar magnetic field & $5\times10^{11}~$G & $\surd$ \\
  $B_{\rm 0max}$\footnote{For the SN link models, this value is the average initial surface magnetic field, $B_{\rm 0av}$, not the maximum} & Initial maximum pulsar magnetic field & $4\times10^{12}~$G & $\surd$ \\
  \hline \hline
  \textsc{binkin} & & &  \\
  \hline \hline
  \multirow{2}{*}{Parameter} & \multirow{2}{*}{Description} & Value/ & Varied in \\
   & & choice & KH09\\
  \hline
  $N$ & Number of systems evolved (also used in \textsc{binpop}) & $10^9$ & $\surd$ \\
  $V_\sigma$ & Maxwellian dispersion for the supernova kick speed (also used in \textsc{binpop}) & $190~$km s$^{-1}$ & $\surd$ \\
  $t_{\rm max}$ & Age of the galaxy (also used in \textsc{binpop}) & $10~$Gyr & $\surd$ \\
  $\Phi$ & Gravitational potential & Pac90 & $\surd$ \\
  $\alpha_{\rm g}$ & Galaxy size and mass scaling parameter normalised to Milky Way & $1$ & $\times$ \\
  $R_{\rm init}$ & Galactic radial stellar birth distribution & YK04 & $\surd$ \\
  $|z_{\rm imax}|$ & Maximum possible birth height off the galactic plane & $0.075~$kpc & $\surd$ \\
  $|z_{\rm max}|$ & Maximum height from the galactic plane used to calculate statistics & $10~$kpc & $\surd$ \\
  \hline
  \end{tabular}
  \end{minipage}
  \end{table*}

\section{Bound double neutron star and black hole population characteristics}
\label{s:DCBNS-NS}

In this section, the population characteristics of 
DCBs are investigated.
%
Model C$^{'''}$ was primarily used by KH09 to
check the accuracy of scale heights 
drawn from models with smaller binary populations. 
The scale height is twice the e-folding distance 
of the population in $|z|$ and is calculated by counting all
systems within $10~$kpc of the plane.
Exploring the DCB scale heights of Model C$^{'''}$, Table~\ref{t:table2} 
shows the clear difference in NS-NS scale heights compared to those 
systems with at least one BH.
This is primarily due to our assumption that BHs do 
not receive any velocity kick during their formation.
It is interesting that BH-NS systems have a greater 
scale height than NS-BH systems.
This results from the order in which the SN kick occurs,
further discussion on this point follows below.
For interest we note that very large distances from the plane
can occur for DCB systems
-- up to $5.7~$Mpc for the NS-NS systems -- indicating
that ejection of DCBs into the intergalactic medium 
can occur.
Such systems are relatively rare, and are those which
receive velocity kicks greater than $\sim 500~$km s$^{-1}$
(and integrated for close to $10~$Gyr).
Very few ($< 1\%$) BH-BH systems receive relatively large
recoil velocities during SN solely from instantaneous mass loss
(see Section~\ref{s:recoilvels}).
Model C$^{'''}$ produces a high relative number of BH-BH systems 
which remain bound after passing through two SNe as 
opposed to systems which receive one or two kicks in 
the process (see also the formation rates calculated in 
Section~\ref{s:fratesNS-NS}), 
once again a result of the assumption in Model C$^{'''}$ that BHs do not receive kicks. 

When BHs are allowed to receive kicks the models show an
expected increase in the scale height of BH-BH DCBs and all systems
that include BHs.
The relative number of BH-BH systems has also decreased,
although they still dominate the DCB numbers.
Such a large increase in scale heights of NS-BH systems, especially
compared to BH-NS systems, is because BH kicks in our model are taken from
a Maxwellian distribution with no regard to the amount of fall-back
onto the BH.
BHs with large mass progenitors ($M>40~M_\odot$) have complete
fall-back; therefore, binary systems are not unbound owing to instantaneous 
SN mass-loss from binary (as there is none) which allows the possibility
of greater SN recoil velocities to be imparted onto the system than otherwise.
Decreasing the efficiency of removing the envelope during the CE phase
significantly reduces the NS-NS scale height.
A lower $\alpha_{\rm CE}$ results in more merger events during CE and
tighter orbits of those systems that do survive.
Although tighter pre-second supernova binary system would allow at first glance
greater recoil velocities to occur (owing to larger allowed SN kicks) 
we find that the magnitude of the recoil velocity after the second SN in NS-NS systems 
is typically smaller ($\sim 200~{\rm km s}^{-1}$) than that in 
Model C$^{'''}$ ($\sim 500~{\rm km s}^{-1}$, see Figure~\ref{f:fig15NS-NS}), 
although the same range is covered. 
NS-NS systems in this model, therefore, typically merge on a smaller time scale than 
otherwise (see Section~\ref{s:coal}).
As expected including EC SN kicks reduces the NS-NS scale height by half, while slightly
decreasing the scale heights of other systems that include a NS.
The relative number of NS-NS systems increases by two orders of magnitude 
in this model. 

\begin{table}
 \centering
 \begin{minipage}{84mm}
  \caption{
    For each model the scale heights (kpc) and relative numbers are provided 
    for all double compact binaries within 
    $10~$kpc of the Galactic plane at a Galactic age of $10~$Gyr.
  \label{t:table2}}
  \begin{tabular}{ccrrrrrrr}
  \hline
 Model & System & Scale height & Relative \\
  & type & [kpc] & number \\
  \hline
 C$^{'''}$ & NS-NS & $1.53$ & $0.003$ \\ 
 & NS-BH & $0.10$ & $0.008$ \\ 
 & BH-NS & $0.24$ & $0.006$ \\ 
 & BH-BH & $0.03$ & $0.982$ \\ 
    $\alpha_{\rm CE} = 1$ & NS-NS & $0.90$ & $0.002$ \\
   & NS-BH & $0.04$ & $0.003$ \\ 
   & BH-NS & $0.15$ & $0.008$ \\
   & BH-BH & $0.03$ & $0.987$ \\
   BH kicks & NS-NS & $1.55$ &  $0.111$ \\
   & NS-BH & $1.18$ & $0.063$ \\
   & BH-NS & $0.46$ &  $0.036$ \\
   & BH-BH & $0.78$ & $0.790$ \\
   BH kicks \& & NS-NS & $0.73$ & $0.290$ \\
   EC SNe & NS-BH & $1.21$ & $0.048$ \\
   & BH-NS & $0.37$ & $0.048$ \\
   & BH-BH & $0.78$ & $0.614$ \\
  \hline
  \end{tabular}
  \end{minipage}
\end{table}

\subsection{Recoil velocities}
\label{s:recoilvels}
\begin{figure}
  \includegraphics[width=84mm]{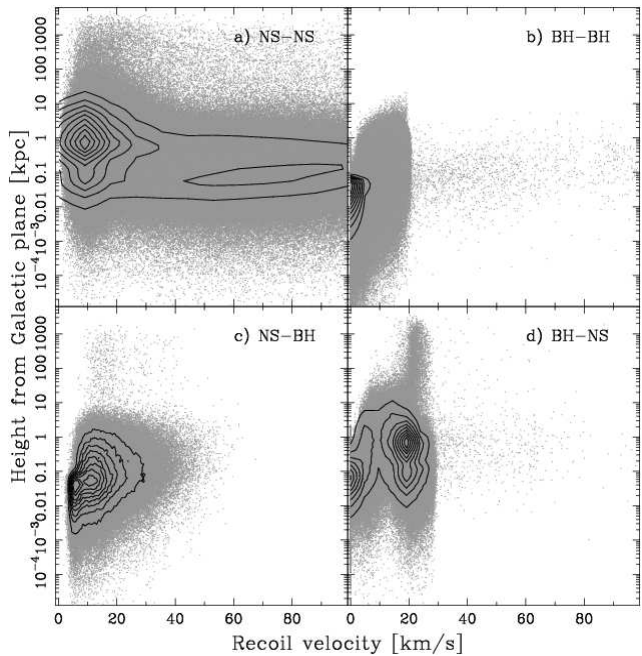}
  \caption{
    Distribution of the centre-of-mass recoil velocities after the 1st
    SN kick and the final system height above the plane 
    in $|z|$ at $10~$Gyr for all double compact systems that formed  
    within Model C$^{'''}$. 
    This includes both coalesced and non-coalesced systems. 
    Provided are contours ranging from $10\%$ through to the
    outer contours of $90\%$.
    Clockwise from the top left is NS-NSs, BH-BHs, BH-NS (BH
    formed first) and NS-BH (NS formed first).
    Because of the large number of BH-BH systems (that may or
    may not coalesce) produced in this model we have
    had to limit the number of points we plot to every $10$th 
    (the contours are calculated from the complete sample).
    \label{f:fig12NS-NS}}
\end{figure}

Figure~\ref{f:fig12NS-NS} depicts the center of mass 
recoil velocities directly after the 
initial SN kick for Model C$^{'''}$.
These are shown as a function of final height
from the Galactic plane for each system at the end of the
simulation (or the point of coalescence).
We compare recoil velocities of the four 
double compact binary types.
We limit the figures to solely those of Model C$^{'''}$,
although our discussion of the resultant populations
cover the other models.
The initial SN velocity kick of those systems 
that form DCBs is typically less than 
the average velocity kick a NS in our models is given.
This was also found by Voss \& Tauris (2003, see Figure 11 
of their paper) and arises because 
typical SN kick velocity magnitudes tend to disrupt 
binary systems, thus preventing DCB formation.
NS-NS progenitors can survive the first SN with 
a greater recoil velocity than their NS-BH 
counterparts (who also receive a kick on the first SN).
The cause of this is partly due to the greater average 
total system mass of NS-BH progenitors compared 
to that of NS-NS progenitors.
However, for the formation of NS-BH binaries, 
the necessity of: mass transfer prior to any SN event, 
mass and orbital angular momentum loss via winds and 
the instantaneous mass loss during SN events (with the 
possibility of fall back onto the stellar remnants), 
greatly complicates matters.
Other than a weak trend for NS-BH systems there does not 
seem to be any significant trend with final system height off the 
plane and the recoil velocity after the first SN.

\begin{figure}
  \includegraphics[width=84mm]{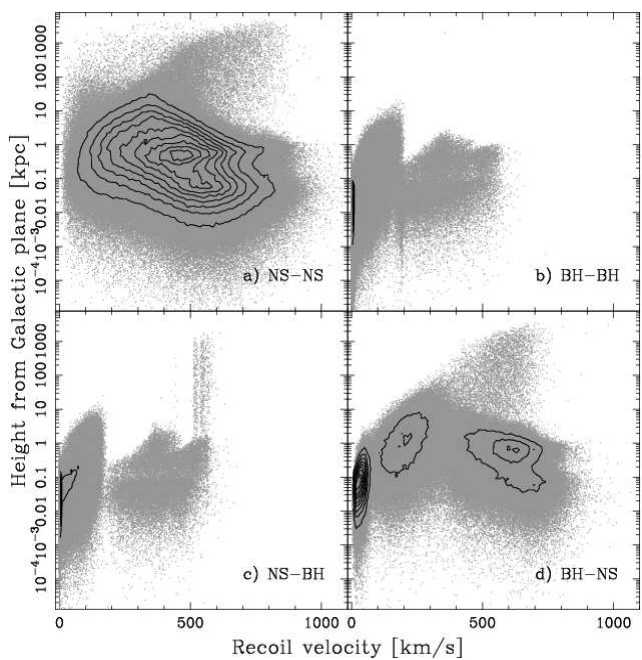}
  \caption{
    As for Figure~\ref{f:fig12NS-NS} but now showing the centre-of-mass recoil velocity 
    induced by the 2nd SN only. 
    Contours are provided as in Figure~\ref{f:fig12NS-NS}.
     \label{f:fig15NS-NS}}
\end{figure}

Figure~\ref{f:fig15NS-NS} illustrates the system centre-of-mass 
recoil velocity induced solely by the second SN.
The range of resultant velocities is much greater 
than that provided by the first SN kick.
Again, like Voss \& Tauris (2003), we find that 
the double compact systems can survive, on average, 
greater second SN kicks than that which may occur for the first SN.
Due to the massive nature of DCB progenitor stars
it is typical that such systems have passed through at 
least one CE phase (see Sections~\ref{s:eccporb} and \ref{s:coal}; 
although see also Dewi, Podsiadlowski \& Sena 2006).
A large fraction of these systems have engaged in CE evolution
in the intervening time between SN events because, when they
survive the CE, the likelihood of binary disassociation
owing to SNe decreases.
Therefore, DCBs are typically tighter prior to the 
second SN compared to the first SN.
Thus the SN kick must overcome greater orbital 
binding energy to disassociate the system which means
a greater SN kick velocity is allowed that may in turn
induce a greater recoil velocity into 
the system (ignoring mass loss).
For some systems the only way that they may survive is
from a sufficiently large and well directed kick,
as discussed in Portegies Zwart \& Yungelson (1998),
and similar to the
low-mass X-ray binary direct formation mechanism
first recognised in Kalogera (1996).
The mass of the double compact progenitors, at the second SN, is 
less than the mass at the first SN and as such the systems
can end with much greater recoil velocities after the second SN.

We note that Figure~\ref{f:fig15NS-NS} contains all systems that
remain bound after the second SN regardless of whether they 
subsequently coalesce or not.
Focusing on the NS-NSs (top left panel of Figure~\ref{f:fig15NS-NS})
we find that of the NS-NS systems in Model C$^{'''}$ that survived 
the two supernovae that $90\%$ coalesce within $10~$Gyr.
Because of this the contours mainly trace the coalesced systems, 
which slope downwards in $|z|$ with increasing recoil velocity.
This inverse trend arises because we simply plot the merger site 
-- the remnant evolution is ignored here -- 
and systems that receive a greater recoil
velocity (on average greater kick velocity and thus 
higher eccentricity) are more likely to merge faster 
than those that receive medium to low recoil velocities.
However, the height off the plane for systems that remain bound and do
not coalesce within our assumed age of the Galaxy is proportional
to the recoil velocity of the second SN (and, of course,
to the age of the system).
This is shown by the population of points in the upper right
of the panel (noting that the trend continues to low
recoil velocity and $|z|$).

The top right panel depicts the BH-BHs.
The majority of systems are contained at small recoil 
velocity values.
However, there is a population of systems with recoil
velocities greater than $\sim 200~$km s$^{-1}$.
These systems are ones that have coalesced within $10~$Gyr
of their formation.
The NS-BH systems are given in the lower left panel
of  Figure~\ref{f:fig15NS-NS}.
Both coalesced and bound NS-BH systems are tightly constrained
to low recoil velocities and both tend to have increased
$|z|$ values with higher recoil velocities, in this low 
recoil velocity range.
However, beyond a recoil velocity of $\sim 200~$km s$^{-1}$ there is
a population of coalescing NS-BH systems which appear
constrained to $|z| < 1~$kpc from the plane of the Galaxy.
There is also a small population of NS-BH systems, which do not coalesce, 
that reside within a narrow recoil velocity band between
$500$ and $600~$km s$^{-1}$.
Such systems extend from $|z| \sim 0.01~$kpc up to $|z| \sim 1000~$kpc.
These systems do not receive SN velocity kicks for the second
SN, so their formation and large $|z|$ does not rely on a sympathetic 
kick direction but instead on 
being tightly bound prior to BH formation and that for 
these systems there has been enough time in their evolution 
to reach such great distances.

The BH-NS systems within the bottom right panel contain
two quite different distributions.
The bound BH-NS systems extend from very low recoil velocities 
and small $|z|$ values to quite large recoil velocities 
and high $|z|$ values.
In contrast the distribution of BH-NSs that coalesce begins 
at recoil velocities of $\sim 200~$km s$^{-1}$ and 
$|z| \sim 1~$kpc and stretches to high recoil velocities 
and a height above the plane range of $0.0001 < |z| < 1~$kpc.
When BHs receive SN kicks their recoil velocity-$|z|$ distribution
takes on the characteristics of the NS counterparts (although with 
generally smaller magnitudes).
Providing NSs with the option to receive lower kick velocities during
EC SNe assembles the resultant NS-NS distribution as an amalgamation 
between all the Model C$^{'''}$ distributions.
When assuming a less efficient CE phase the NS-NS systems have a 
bimodal recoil velocity distribution with a group clustered between velocities of $0~{\rm km s}^{-1}$ 
to $\sim 30~{\rm km s}^{-1}$ and another centered around $\sim 100~{\rm km s}^{-1}$
and extending either side by roughly $40~{\rm km s}^{-1}$. 

\subsection{Eccentricities and orbital periods}
\label{s:eccporb}

\begin{figure}
  \includegraphics[width=84mm]{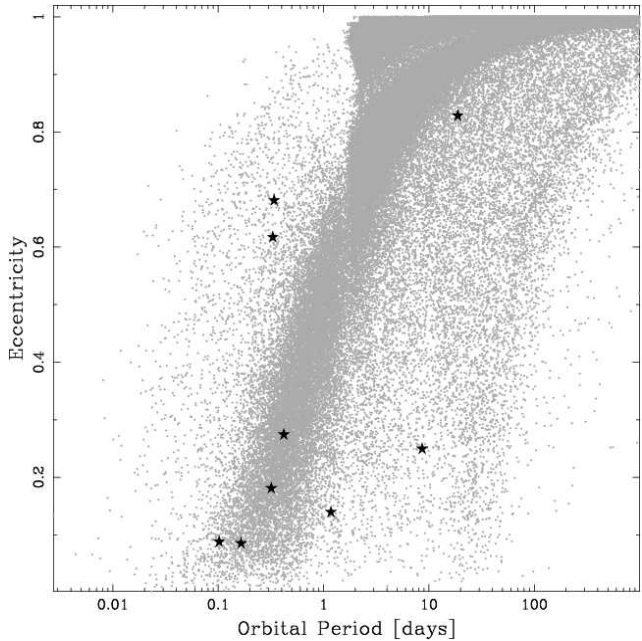}
  \caption{
    Eccentricity and orbital period parameter space
    snapshot of NS-NS systems at a Galactic age of $10~$Gyr
    and within $4.5 < R < 12.5~$kpc.
    Included over our model points are $8$ observed
    systems suspected to be NS-NSs according to 
    van den Heuvel (2007) and Stairs (2008). 
    The observed systems are,
    J0737-3039 (Lyne et al. 2004), 
    J1518+4904 (Nice et al. 1996), 
    B1534+12, J1811-1736, B1913+16 (Stairs 2004), 
    J1756-2251 (Faulkner et al. 2005),
    J1829+2456 (Champion et al. 2004),
    J1906+0746 (Lorimer et al. 2006b) and B2127+11C (Anderson et al. 1990;
    detected within M15 with an eccentricity of 0.68 and orbital
    period of 8.05 hours).
    \label{f:fig222NS-NS}}
\end{figure}

To examine NS-NS systems in greater detail 
Figure~\ref{f:fig222NS-NS} depicts the 
eccentricity-orbital period population distribution
that occurs at the simulation end.
With greater numbers of NS-NSs now known (9 systems, 8 within
the Galactic disk) and more realistic  mass estimates of 
their stellar components available, recently it has 
become clear that this diagram sheds light on a 
possible inconsistency between observations and stellar 
binary theory (van den Heuvel 2007).
Similar distributions to those shown in Figure~\ref{f:fig222NS-NS} 
have been discussed in varying detail
previously in population synthesis works 
(Portegies Zwart \& Yungelson 1998;
Bloom, Sigurdsson \& Pols 1999; Voss \& Tauris 2003; 
Ihm, Kalogera \& Belczynski 2006).
Unlike these previous works the systems considered here
are restricted to reside within the solar neighbourhood 
($4.5 \geq R \geq 12.5~$kpc).
There is also no assumption on the ages of the 
observed NS-NS systems used for comparison -- that is, the
complete randomly born NS-NS model population is shown.
Like Portegies Zwart \& Yungelson (1998) this model
finds that the addition of SN kicks within NS formation
smears the eccentricity and orbital period distributions.
Portegies Zwart \& Yungelson (1998) also detail the 
importance of CE evolution in shaping the 
eccentricity-orbital period distribution.
The typical formation pathway of NS-NSs systems that reside in
Figure~\ref{f:fig222NS-NS} pass through stable
mass-transfer prior to any SN event.
Owing to this mass-transfer either the
mass ratio $q$ inverts or, if $q \sim 1$, the 
secondary stars (initially less massive accreting stars)
grow significantly more massive driving $q$ to
small values.
After the first SN event the systems have a large 
range of eccentricities and orbital periods, however,
with the onset of secondary star evolution 
these systems eventually pass through a CE evolutionary 
phase, circularising the orbit.
The binary systems must survive CE and the secondary stars
evolve to explode, forming NSs.
It is the orbital parameters at the time of this SN 
and the event (kick) itself which is important in 
regulating the resultant eccentricity-orbital 
period NS-NS population distribution.
Such an evolutionary pathway compares well with the models
of Portegies Zwart \& Yungelson (1998), though, for 
young NS-NSs Portegies Zwart \& Yungelson find many systems
that pass through multiple CE phases between the two SN events.
Multiple CE systems are also found in this model, however,
these systems coalesce rapidly (few Myr), as such a
discussion of these systems is left until Section~\ref{s:coal}.

The distribution of NS-NS systems within Figure~\ref{f:fig222NS-NS}
covers well the nine observed systems 
believed to be NS-NSs (van den Heuvel 2007; Stairs 2008).
However, there is a greater relative 
number of observed NS-NSs with low eccentricities than 
that produced in this model.
The majority of modelled systems within Figure~\ref{f:fig222NS-NS}
have eccentricities greater than $0.7$ and only $17\%$ have $e < 0.5$.
Of the observed Galactic disk NS-NS systems $75\%$ have $e < 0.5$.
This discrepancy may be due to low number statistics
owing to pulsar observational selection effects or incomplete theoretical 
modelling. 
One possible evolutionary phase that may assist in lowering the
average eccentricity of model NS-NSs, as suggested by 
van den Heuvel (2007), is electron capture SNe (Miyaji et al. 1980; 
and as discussed in Section~\ref{s:bpbk}).
EC SNe would induce smaller SN kicks into these systems 
potentially rendering them less eccentric.
We are now in a position to examine this possibility with our
EC SN model.
We find that with EC SN included the percentage of systems with
$e < 0.5$ increases only slightly to $20\%$.
The overabundance of highly eccentric model NS-NS systems
still remains a problem.
We note here that a full parameter space search that includes EC SNe kicks
can produce models that match the observed eccentricity distribution
(Andrews, Kalogera \& Belczynski, in preparation).
However, we show here that this is not the case for standard model 
assumptions, of typical CE efficiencies and core collapse SN velocity kicks.

\subsection{Eccentricities and height above the plane}

\begin{figure}
  \includegraphics[width=84mm]{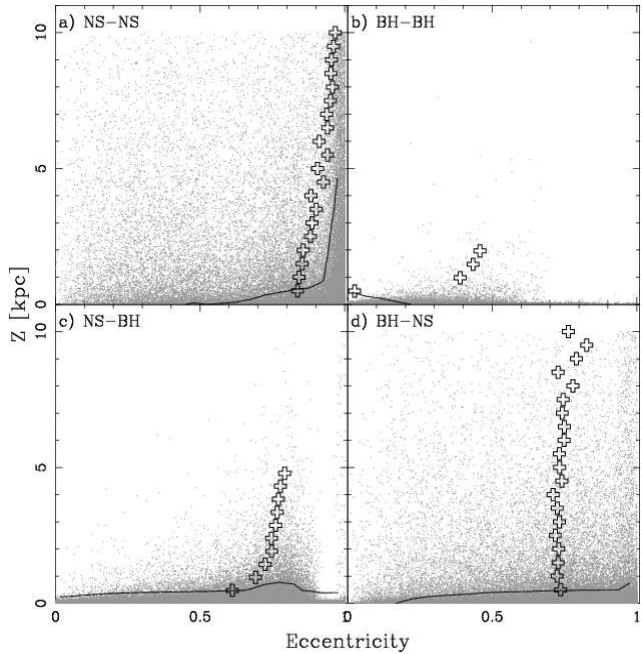}
  \caption{
    The eccentricity-$|z|$ distributions for each 
    double compact system type.
    Top left is NS-NSs, top right BH-BHs, bottom left NS-BH
    where the NS formed first and bottom right BH-NS
    where the NS formed second.
    The $90\%$ contour is shown for guidance.
    We also provide a representation of the median 
    eccentricity of the 
    distribution in steps of $|z| = 0.5~$kpc (plus symbols).
    When the statistical significance of each median 
    eccentricity value is poor (less than $10$ systems
    in the $|z|$ bin) we do not plot the point.
    \label{f:fig16NS-NS}}
\end{figure}

It is also possible to examine the Galactic dynamics of 
double compact binaries to look for correlations 
between orbital properties and location.
For such an analysis we make use of Figure~\ref{f:fig16NS-NS} which
depicts the $|z|$-eccentricity parameter space for the four
double compact binary types.
The kinematics of NS-NS systems is as expected: the large 
relative number of eccentric systems typically extend in 
greater numbers further out in $|z|$ than less eccentric 
systems.
The contour shows a strong trend within $1~$kpc
of increasing eccentricity with increasing $|z|$.
However, there is effectively no variation above $1~$kpc, 
the median eccentricity does not vary significantly with 
increasing height above the plane (see plus symbols within
Figure~\ref{f:fig16NS-NS}).
This is not the case for BH-BH systems of which the majority
reside close to the Galactic plane with little to no eccentricity.
Here the median eccentricity increases with rising $|z|$ until
low number statistics dominate at $|z| \sim 2~$kpc.
This suggests that the recoil velocity after the SN is greater
with greater induced eccentricity.
Because BH-BH systems receive no SN velocity kick, 
the resultant eccentricity and centre-of-mass recoil velocity 
relies upon the initial orbital stellar velocity 
components and the instantaneous mass loss from the system.
According to the SN fall back prescription used within
Kiel et al. (2008; from Belczynski, Kalogera \& Bulik 2002) 
the greater the SN progenitor the less
mass is lost from the system during SN and for many massive BHs
complete fall back occurs.
Therefore, in some instances there is little to no
recoil velocity induced onto the system. 
Similar to BH-BH systems, although shifted to 
higher eccentricities, is the $|z|$-eccentricity 
trend of NS-BH systems.
There are more eccentric binaries in this population than 
in the BH-BH population.
Owing to the kick velocity imparted during the NS formation,
many systems are still eccentric when the BH is formed.
The eccentricity ratio trend of the NS-BH systems, in $|z|$, 
is similar to that of the BH-BH population because the 
final SN produces a BH -- the higher recoil velocity
is directly proportional to the induced eccentricity.
This is why there is no such trend in the BH-NS population: 
the eccentricity-recoil velocity link is weakened by the
SN kick (the eccentricity depends mostly upon the kick strength, the
recoil velocity upon the kick strength, direction and orbital
parameters).

Because distributions such as in Figure~\ref{f:fig16NS-NS} 
rely on SN kicks and binary evolution further observations 
of pulsar populations measuring parameters such as
eccentricity, orbital period and height from the Galactic
plane may help to constrain these evolutionary
features.
Of the 8 observed Galactic disk NS-NS systems all reside within 
$|z| < 1~$kpc and have a median eccentricity of 
$0.2$.
This is significantly less than the medium eccentricity produced in our model.
Of the observed systems four reside within $0.05~$kpc of the
plane and while we do find that the median eccentricity
of the model NS-NSs decreases with $|z|$ (for $|z| < 1~$kpc)
we do not get below a value of $0.5$.
As shown in our analysis of Figure~\ref{f:fig222NS-NS} 
this discrepancy is not resolved when accounting for core collapse 
electron capture.
Even limiting the region of the Galaxy considered to $|z| < 1~{\rm kpc}$
and $4.5 < R < 2.5~{\rm kpc}$ does not rectify the situation,
only increasing the percentage of systems with $e < 0.5$ to $\sim 25\%$.
Of course observational selection effects occur and unless 
these are also accounted for any direct observational 
comparisons are incomplete.\footnote{Accounting for selection effects
will be the focus of follow-up work (Kiel, Bailes \& Hurley, in preparation)}

\subsection{Orbital period and Galactic kinematics}

\begin{figure*}
  \includegraphics[width=168mm]{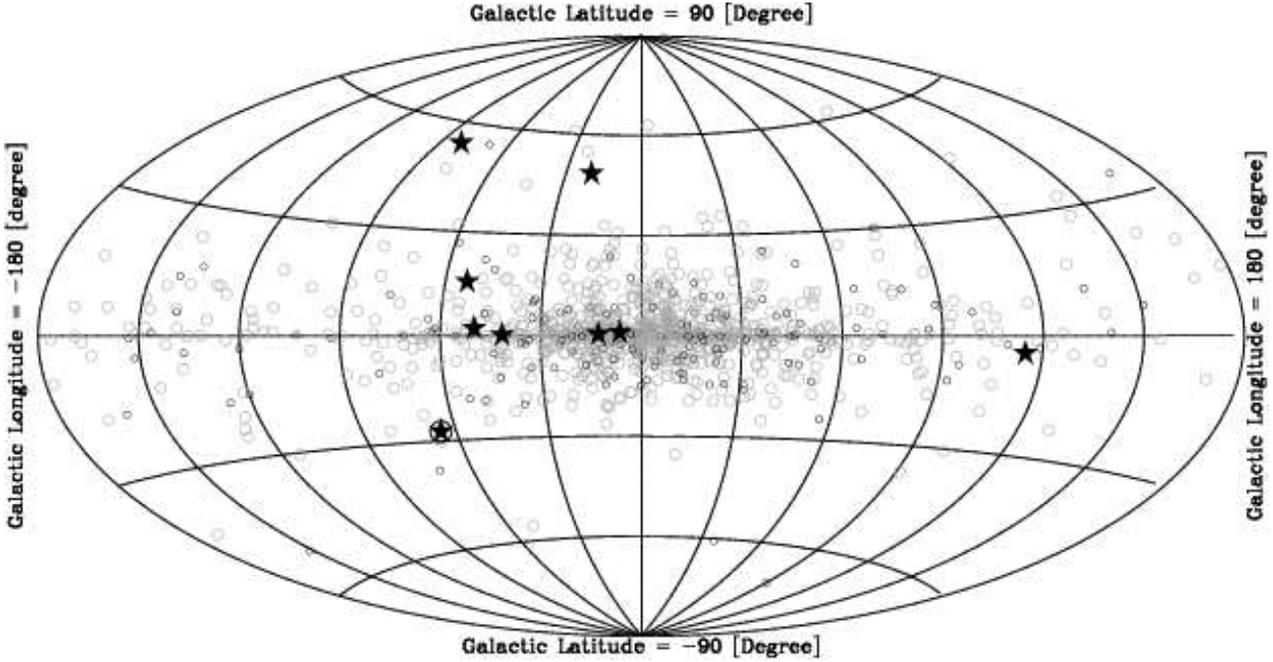}
  \caption{
    Aitoff projection of a subset of NS-NSs (circles)
    within $|z| < 5~$kpc of the Galactic plane
    and $R < 30~$kpc of the Galactic centre.
    The NS-NS systems shown all have $P_{\rm orb} < 1~$day.
    Black circles represent systems with eccentricity
    greater than $0.5$, the grey less than $0.5$.
    Here a longitude of zero is the Galactic centre.
    For simplicity this figure has been made from Model C,
    that is, the number of underlying systems used to produce
    this plot is a factor of a $100$ less than in Model C$^{'''}$.
    We include the 9 detected NS-NS systems as black stars
    as found in the ATNF Pulsar Catalogue (the detected system
    enclosed in a large black circle is B2127+11C).
    \label{f:fig210NS-NS}}
\end{figure*}
It is informative to examine the distribution of NS-NS systems
in Galactic coordinates, in particular, to investigate the
optimal regions within the Galaxy to survey when searching for
such systems.
Figure~\ref{f:fig210NS-NS} depicts this for those NS-NSs with orbital
periods less than $1~$day in an Aitoff projection (as designed by
Hammer 1892: see Steers 1970). 
We provide two populations:
those with eccentricities less than $0.5~$ (grey circles)
and those with eccentricities greater than $0.5~$ (dark circles).
NS-NS systems, according to our model, will most likely be observed 
towards the Galactic centre.
In Galactic Cartesian coordinates the distribution in $R$
peaks at $R \sim 5~$kpc and the systems preferentially 
reside close to the Galactic plane. 
Unfortunately for observational predictions the populations
with low and high eccentricities trace each other quite well.
We provide the 9 detected NS-NS systems as a guide.

\section{Coalescing double compact binaries}
\label{s:coal}

We now examine the population of double compact systems
that coalesce within the simulations (i.e. within the assumed 
age of the Galaxy).
The scale heights of these merger events are given in 
Table~\ref{t:table3}.
The most interesting aspect of Table~\ref{t:table3} is the
scale height of coalescing NS-NSs compared to coalescing BH-NSs.
It is surprising that the systems which typically receive
greater combined recoil velocities (from both SN events)
have a lower scale height.
This scale height difference does not arise from low number 
statistics but rather from the shorter merger time scales
of NS-NS systems compared to BH-NSs.
Once a close NS-NS is formed, if it is to coalesce within 
a Hubble time, it will generally coalesce after
a few Myr (see also Chaurasia \& Bailes 2005).
Whereas after BH-NS formation coalescence
typically occurs after a few Gyr.
Of course, this story changes somewhat when we include
EC SN into the models.
Now the scale height for merging double NSs increases
owing to an increase in the typical merger time scale.
Including kicks to BHs obviously helps to increase 
the scale height and decrease the merger time scale.
Assuming $\alpha_{\rm CE} = 1$ causes NS-NS systems to be
closer after the final supernova explosion, which allows these 
systems to merge faster and decreases their merger scale height.

\begin{table}
 \centering
 \begin{minipage}{84mm}
  \caption{
    Model scale heights (kpc) and relative numbers for all 
    DCBs that have coalesced for the four models.
  \label{t:table3}}
  \begin{tabular}{ccrrrrrrr}
  \hline
 Model & System & scale height & Relative \\
  & type & [kpc] & number \\
  \hline
  C$^{'''}$ & NS-NS & $0.51$ & $0.239$ \\ 
   & NS-BH & $0.11$ & $0.029$ \\ 
   & BH-NS & $0.69$ & $0.041$ \\ 
   & BH-BH & $0.05$ & $0.691$ \\ 
   $\alpha_{\rm CE} = 1$ & NS-NS & $0.40$ & $0.051$ & \\
   & NS-BH & $0.06$ & $0.001$ \\ 
   & BH-NS & $0.05$ & $0.055$ \\
   & BH-BH & $0.05$ & $0.893$ \\
   BH kicks & NS-NS & $0.53$ & $0.536$ \\
   & NS-BH & $1.06$ & $0.049$ \\
   & BH-NS & $0.69$ & $0.029$ \\
   & BH-BH & $0.66$ & $0.386$ \\
   BH kicks \& & NS-NS & $0.82$ &$0.827$ \\
   EC SNe & NS-BH & $1.07$ & $0.018$ \\
   & BH-NS & $0.67$ & $0.011$ \\
   & BH-BH & $0.65$ & $0.144$ \\
  \hline
  \end{tabular}
  \end{minipage}
\end{table}

\begin{figure}
  \includegraphics[width=84mm]{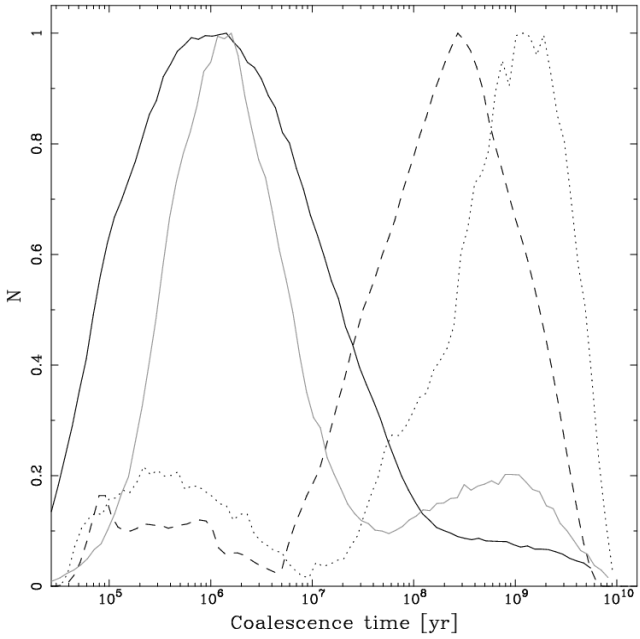}
  \caption{
    The coalescence times after DCB formation for Model C$^{'''}$
    measured as the time elapsed between
    DCB formation and merger of the two stars.
    Full black line gives NS-NSs, dashed line is BH-BHs, 
    full grey line is NS-BHs while the dotted line is BH-NSs.
    The distribution for each type of system is normalised to unity.
    This figure only counts those DCBs that coalesced within our simulation,
    which introduces a selection effect at large coalescence times. 
    \label{f:fig182NS-NS}}
\end{figure}

\begin{figure}
  \includegraphics[width=84mm]{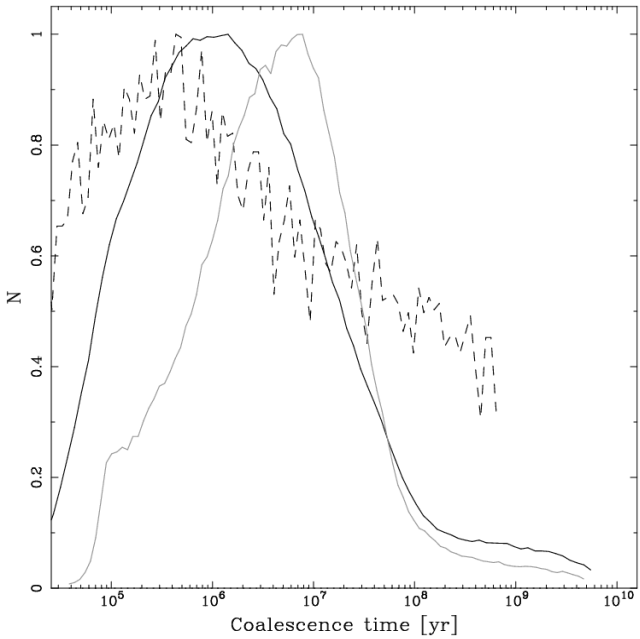}
  \caption{
    The coalescence times after NS-NS formation for
    variants on Model C$^{'''}$.
    The full black line depicts Model C$^{'''}$ (as in Fig.~\ref{f:fig182NS-NS}).
    The model including electron capture SNe is given by the grey line and
    the dashed line depicts the model with $\alpha_{\rm CE} = 1$.
    Each distribution is normalised to unity.
    \label{f:NS-NSmergertime}}
\end{figure}

\begin{figure}
  \includegraphics[width=84mm]{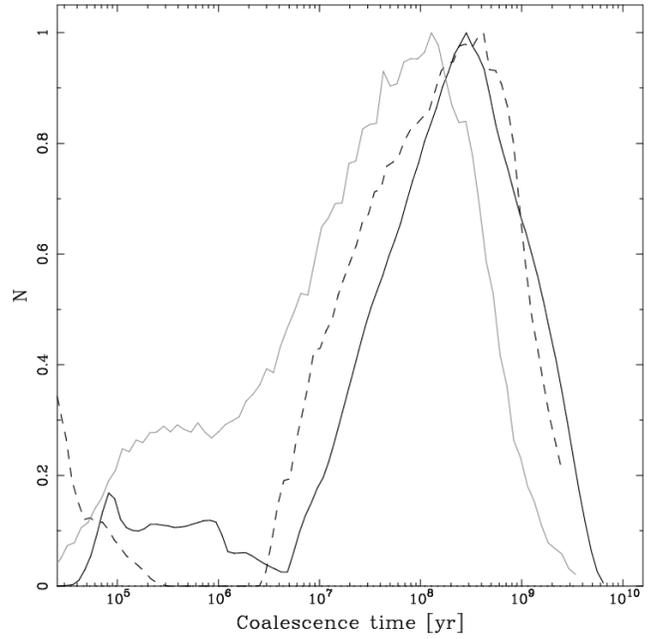}
  \caption{
    The coalescence times after BH-BH formation for
    variants on Model C$^{'''}$.
    The full black line depicts Model C$^{'''}$ (corresponding to the dashed line 
    in Fig.~\ref{f:fig182NS-NS}). The model including 
    SN velocity kicks for BHs is denoted by the grey line and
    the dashed line depicts the model with $\alpha_{\rm CE} = 1$.
    Each distribution is normalised to unity.
    \label{f:BH-BHmergertime}}
\end{figure}

The coalescence times for the four system types in Model C$^{'''}$ are 
given in Figure~\ref{f:fig182NS-NS} with each distribution 
normalised to unity. 
The coalescence times shown in Figures~\ref{f:fig182NS-NS} to~\ref{f:BH-BHmergertime}
are for those systems that merged during the simulation and represent the time each system 
took to coalesce (the length of time the DCBs lived). 
We note that there are in fact many more
BH-BH and NS-NS systems than NS-BH and BH-NSs 
(an order of magnitude) and also we do not include the projected
coalescence times for these systems that do not merge within our model.
A model selection effect is then introduced at large times, where a
turn over in the curves occur.
Much can be gleaned from the incidence of the merger 
timescales peaks between each system type.
The peaks of NS-NS and NS-BH systems correlate well, 
as do the peaks for BH-BH and BH-NSs.
From this alone we see the importance of the first
SN on the final system merger timescale.
The systems in which a NS forms from the first SN --
imparting an asymmetric kick into the system --
typically merge faster than those in which a BH is
formed from the first SN.
Also those systems that pass through multiple CEs merge fastest.
In DCB formation mass transfer prior to any SN is 
an important factor in determining the number of CE 
events beyond the first SN event.
If unstable mass transfer occurs prior to any SN
and a CE occurs it is unlikely that two CE events will
occur following the first SN event.
However, if stable mass transfer occurs prior to the
first SN it is possible that if the orbit is wide enough 
after the first SN then the companion star will initiate
unstable mass transfer, while, say, core helium burning
takes place, leaving a tight enough orbit for the resultant
helium star to evolve and overflow its Roche-lobe and
cause unstable mass transfer.
Tightening the orbit once more.
This is also more likely to occur -- in a relative sense --
for NS-NSs and NS-BHs than for BH-BHs and BH-NSs, because the
NS SN kick can cause a tight binary system to expand 
more readily than for a BH SN event.
These systems (NS-NSs and NS-BHs) also have a high 
eccentricity which allows
unstable mass transfer to occur even though the separation
of the two stars would otherwise be greater.
The BH-BH and BH-NS systems that are wide (tens of 
thousands of solar radii) after the first SN
must be wide to begin with.
Unfortunately for these systems the relative stellar
velocities are small which means that little to no
eccentricities are imparted to the orbit.
If after the first SN the secondary initiates mass transfer
while core helium burning is taking place then, with
such little eccentricities and large separations, these
systems are likely to have stable mass transfer and 
expand their orbits.
Even if a CE phase does occur, after
spiral in, the systems are still hundreds of 
solar radii apart.

The NS-NS and NS-BH systems that merge quickly all survive
the first CE phase with a separation of $1-2~R_\odot$.
The helium giant secondary star may then evolve to initiate another
CE phase, in which the second NS is formed.
Again the system survives with a separation
of $1-2~R_\odot$.
Such short orbital periods allows rapid (few Myr)
coalescence.
BH-BH and BH-NS systems that merge within $10~$Gyr
generally survive the CE phase with separations
greater than $5~R_\odot$.
The secondary star evolves and forms either a BH or NS
but the separation at formation is typically of order
$10~R_\odot$, for which a system will coalesce on a
Gyr timescale.
Therefore, these systems are susceptible to changes in 
the assumptions of CE evolution.
However, usually such changes only result in shifts 
in the progenitor masses and orbital periods that 
produce such systems -- the evolution pathway forming each 
system type remains viable (Kiel \& Hurley 2006) but the relative
numbers between system types change.
Population birth/death rates (an important tool in 
model comparisons to observations) on the other hand 
are sensitive to changes in progenitor properties 
(Belczynski, Kalogera \& Bulik 2002; O'Shaughnessy et al. 
2005a).
In particular rates are sensitive to changes in the initial stellar mass 
ranges and distributions and orbital period distributions (Kiel \& Hurley 2006). 

The asymmetric kicks in NS-NS and NS-BH systems also help
in forming both very close and disrupted systems.
This is indirectly shown by the greater width of 
the NS-NS merger time peak in Figure~\ref{f:fig182NS-NS} 
to that of the NS-BH peak.
The final NS-NS system separation depends upon the 
second SN kick, which is a random distribution,
and causes a large array of separation values, 
which in turn forces variation into the coalescence
times -- smearing the merger time peak.
The BH-BH merger time scale peak is not quite aligned 
to the BH-NS peak.
Basically, this arises owing to the nature of 
gravitationally induced coalescence: heavier systems
of the same separation after the second SN will
naturally merger faster -- even if the orbits 
are more circular.
Of course, there is some cross over with system types
and the mass transfer phases that occur which is why
the four merger time curves have multiple peaks.
However, the above analysis examines the \textit{main} 
formation pathway for each system type.

We note here that this picture changes somewhat 
when the random birth age of 
each system is accounted for. 
In this case the number of merging systems steadily increases 
over time, 
reaching a maximum at the end of our simulation. 
Take NS-NS systems for example where the 
typical system age at the time of coalescence 
in Model C$^{'''}$ is $20-100~$Myr.
Therefore, with random birth ages included the coalescence
times for the model Galaxy begin at $\sim 10^7~$year. 
From that point on the continuous birth of systems between 
$0-10~$Gyr means that the number of coalescing systems 
increases until the end of star formation.

Of the NS-NS systems \textit{observed} in the Galaxy approximately half
are expected to merge within $10~$Gyr -- based on calculating
the merger time scale owing to gravitational radiation from
their orbital parameters.
The double pulsar J0737-3039 has the shortest measured merger 
time scale of $85~$Myr (Burgay et al. 2003).
Therefore, the merger times of the observed NS-NSs all
fall within the tail of our model distribution.
This means that the model predicts that there is a 
large population of unobserved NS-NSs with merger time
scales shorter than those already observed.
Future detection of such a system has the potential
to significantly increase the empirical Galactic NS-NS 
merger rate (Kalogera et al. 2007).

When accounting for other evolutionary scenarios and assumptions
the merger times of systems can change.
Figures~\ref{f:NS-NSmergertime} and~\ref{f:BH-BHmergertime} depict 
the changes in merger times of NS-NS and BH-BH systems respectively, 
for a variety of models.
When including EC SN evolution into our model the typical time NS-NS systems 
take to merge increases, as denoted by comparison of the peaks between the black
line (Model C$^{'''}$) and the grey line (model including EC SN) within 
Figure~\ref{f:NS-NSmergertime}.
Such an outcome was expected because the small merger timescales of Model C$^{'''}$
are driven by strong well directed kicks that force the orbital separation to contract.
Decreasing the common envelope efficiency from $3$ to unity decreases the number
of DCBs produced, which explains the erratic nature of the dashed curve within 
Figure~\ref{f:NS-NSmergertime}.
The $\alpha_{\rm CE} = 1$ curve peaks at a slightly lower coalescence time than 
the other two models depicted.
The less efficient the CE phase the greater the number of systems that merge or 
survive with smaller orbital separations.
This last point means that systems which go on to form DCBs typically merge faster.
There are, however, systems that normally  would not have survived the second SN
which now do (owing to the stronger gravitational well the exploding star resides in).
There are a greater relative number of these large separation NS-NS systems in lower
CE efficiency models, which can also be seen at long coalescence times in 
Figure~\ref{f:NS-NSmergertime}.
Note that the curves in Figures~\ref{f:fig182NS-NS}, \ref{f:NS-NSmergertime} and 
\ref{f:BH-BHmergertime} end prior to the age of the Galaxy because the number of 
systems drop to extremely low values.
There are DCB populations with longer lifetimes than these end points suggest, however,
the lifetimes are much to long to allow coalescence during our simulations (the DCB population
characteristics are similar to the MSP-BH population described in detail within Section 4.4.2 of
KH09).
Figure~\ref{f:BH-BHmergertime} depicts BH-BH coalescence times for Model C$^{'''}$
(black line), BHs receiving kicks (grey line) and the $\alpha_{\rm CE} = 1$ (dashed line) 
models.
When SN kicks are introduced into BH evolution BH-BH systems merge slightly
faster than otherwise.
When $\alpha_{\rm CE} = 1$ the coalescence time peaks at a similar value to that
of Model C$^{'''}$ but there is a greater relative number of BH-BH systems that merge very
quickly ($< 0.1~{\rm Myr}$).

\subsection{Formation and merger rates}
\label{s:fratesNS-NS}


\begin{table}
 \centering
 \begin{minipage}{84mm}
  \caption{
   Formation rates for double compact binaries in a range of models 
   over a span of $10\,$Gyr. 
   Models considered are Model C$^{'''}$ from KH09 and three variations 
   on this model which are, in turn, a reduction of $\alpha_{\rm CE}$, 
   the inclusion of BH velocity kicks, and the inclusion of BH kicks as well 
   as EC SNe. 
  \label{t:tableFormRates}}
  \begin{tabular}{crrrrr}
  \hline
  & Model & C$^{'''}$ & $\alpha_{\rm CE} = 1$ & BH kicks & BH kicks $+$ \\
  & & & & & EC SNe \\
  Type & & ${\rm Myr}^{-1}$ & ${\rm Myr}^{-1}$ & ${\rm Myr}^{-1}$ & ${\rm Myr}^{-1}$ \\
  \hline
  NS-NS & & $38$ & $4$ & $37$ & $162$ \\
  NS-BH & & $10$ & $2$ & $4$ & $5$ \\
  BH-NS & & $10$ & $7$ & $3$ & $3$ \\
  BH-BH & & $820$ & $750$ & $42$ & $45$ \\
  \hline
  \end{tabular}
  \end{minipage}
\end{table}

\begin{table}
 \centering
 \begin{minipage}{84mm}
  \caption{
   As for Table~\ref{t:tableFormRates} but now showing merger rates for DCBs. 
  \label{t:tableMergeRates}}
  \begin{tabular}{crrrrr}
  \hline
  & Model & C$^{'''}$ & $\alpha_{\rm CE} = 1$ & BH kicks & BH kicks $+$ \\
  & & & & & EC SNe\\
  Type & & ${\rm Myr}^{-1}$ & ${\rm Myr}^{-1}$ & ${\rm Myr}^{-1}$ & ${\rm Myr}^{-1}$ \\
  \hline
  NS-NS & & $36$ & $3$ & $35$ & $154$ \\
  NS-BH & & $4$ & $0.04$ & $3$ & $3$ \\
  BH-NS & & $6$ & $3$ & $2$ & $2$ \\
  BH-BH & & $107$ & $56$ & $25$ & $27$ \\
  \hline
  \end{tabular}
  \end{minipage}
\end{table}

In Tables~\ref{t:tableFormRates} and~\ref{t:tableMergeRates} we show the formation and merger 
rates respectively for the four DCB system types in our models. 
To calculate the rates we first count the fraction of stars in the model that led to 
a Type II supernova, combine this with the assumption that the fraction of binaries 
in the Galaxy is 0.5, and normalize this to the empirical 
Galactic type II SNe rate ($\sim 0.01 \, {\rm yr}^{-1}$: Cappellaro, Evans \& Turatto 1999). 
The relative numbers of the DCB systems and mergers are then calibrated against 
this to produce the rates. 
This method of calculation is commonly employed in binary population synthesis 
(Belczynski, Kalogera \& Bulik 2002; Belczynski, Bulik \& Rudak 2002; 
Voss \& Tauris 2003; Pfahl et al. 2005, for example) 
so facilitates easy comparison with previous work. 
An alternative method uses the Galactic star formation rate (e.g. Kiel \& Hurley 2006). 
With either method the uncertainties involved in the calibration make it more 
appropriate to discuss relative rather than absolute rates. 
This is also true of the uncertainties in rates owing to variations 
in the parameters of binary evolution models. 
These uncertainties have been well documented in the past. 
For example, Belczynski, Bulik \& Rudak (2002) find NS-NS merger rates 
in the range $3$--$300\,$Myr$^{-1}$ as a result of a comprehensive exploration 
of the available parameter space. 
While O'Shaughnessy et al. (2005b) favor models with NS-NS merger rates within the range
$2.5$--$25\,$Myr$^{-1}$, although the total model range of merger rates extends 
far beyond these limits.
Conversely, we find that rates quoted for a particular model (with a set choice of 
parameter values) vary by only a few percent at most with repeated realisations 
using distinct random number distributions (as was also found by 
Belczynski, Bulik \& Rudak 2002). 
We note that in calculating our rates we have assumed an age of $10\,$Gyr 
for the Galaxy. 
If we instead take an age of $15\,$Gyr the quoted rates increase by less than 10\%. 

We quote rates for our Model C$^{'''}$ -- with parameter values and distribution functions 
as listed in Table~\ref{t:binpopbinkin} -- and a set of comparison models in which we 
alter the common-envelope efficiency parameter, the treatment of BH velocity kicks, 
and the possibility of NS formation via electron-capture SNe. 
A number of trends are immediately evident from 
Tables~\ref{t:tableFormRates} and~\ref{t:tableMergeRates}: 
(i) decreasing $\alpha_{\rm CE}$ can reduce the rates by as much as an order 
of magnitude; 
(ii) without kicks for BHs the BH-BH rates are very high; and, 
(iii) the inclusion of EC SNe leads to a sharp increase in the NS-NS rates. 
Looking first at the NS-NS systems the merger rates of all models are in agreement 
with the empirical estimates of  $3$--$190\,$Myr$^{-1}$ made recently by 
Kim, Kalogera \& Lorimer (2006). 
They are also in agreement with the ranges found in the binary population synthesis 
works of Belczynski, Bulik \& Rudak (2002) and O'Shaughnessy et al. (2005b), as is the 
noted behaviour of rates with variations in $\alpha_{\rm CE}$. 
The NS-NS merger rate found by Belczynski, Bulik \& Rudak (2002) in their 
standard model was $53\,$Myr$^{-1}$ which is higher than in our Model C$^{'''}$ 
which has a rate of $36\,$Myr$^{-1}$. 
Belczynski, Bulik \& Rudak (2002) used 
$\alpha^{'}_{\rm CE} \lambda = 1$ in their standard model which, 
assuming $\lambda = 0.5$, means  $\alpha^{'}_{\rm CE} = 2$ and consequently 
a more efficient CE process than in Model C$^{'''}$ (which effectively uses 
$\alpha^{'}_{\rm CE} \lambda \simeq 1$: see Section~2). 
Thus the difference is consistent with the expected trend in CE efficiency but 
there are numerous more subtle differences between the models which could 
also play a role. 


A noticeable outcome of our models is the high formation and merger rates 
for BH-BH binaries predicted by Model C$^{'''}$, 
which are above previous predictions in the literature 
(e.g. Lipunov et al 1997; 
Portegies Zwart \& Yungleson 1998; Voss \& Tauris 2003; 
Belczynski et al. 2007a), in some cases by more than an order of magnitude. 
Following the methodology of Belczynski et al. (2007a) the BH-BH merger rate 
of Model C$^{'''}$, given in Table~\ref{t:tableMergeRates}, sets a
detection rate for the current LIGO gravitational wave detector 
at approximately $1~{\rm yr}^{-1}$.
LIGO has been running in detection mode for longer than this without
a detection and therefore the Model C$^{'''}$ rate is inconsistent and thus an 
over-estimate. 
Reducing the efficiency of the CE spiral-in process halves the predicted BH-BH 
merger rate which is a step in the right direction. 
A further reduction result from the introduction of BH velocity kicks. 
In our model where BHs are given velocity kicks from the same distribution
as for NSs (Maxwellian distribution with dispersion of 
$190~$km s$^{-1}$) the BH-BH formation and merger rates decrease to 
$42~{\rm Myr}^{-1}~{\rm and}~25~$Myr$^{-1}$, respectively. 
We now find that the merger rate of NS-NS systems is larger than for
BH-BH systems in agreement with the majority of previous works
(other than Voss \& Tauris 2003). 
This is further accentuated with the inclusion of EC SNe which increases 
the NS-NS merger rate to $154$ Myr$^{-1}$ while leaving the BH-BH rate untouched. 
The models with BH velocity kicks lead to more plausible merger rate predictions 
as they are safely below the limit requiring a LIGO detection. 
They also agree well with the rates from the models of Belczynski, Kalogera \& Bulik (2002) 
with similar input parameters and assumptions (see Belczynski et al. 2007a for a discussion 
of these rates). 

As mentioned in Section~2, Belczynski et al. (2007a) describe a radically 
different evolutionary pathway to that presented in our models
for Hertzsprung Gap stars which initiate CE. 
That is, these systems never survive CE.
Belczynski et al. (2007a) show that adopting this pathway reduces the merger 
rates of BH-BH systems by a factor of 500 compared to a reduction of 
only a factor of 5 for NS-NS mergers. 
This swing would account for the stark difference in the BH-BH/NS-NS 
merger ratios found in our models compared to 
Belczynski et al. (2007a) and also makes a BH-BH gravitational wave 
detection less likely. 

The detection of gravitational waves with
Advanced LIGO (Lazzarini 2007) will go a long way towards constraining 
the NS-NS and BH-BH merger rates and importantly reduce the uncertainty 
in the BH-BH merger rates predicted by models. 
This has been highlighted previously by Belczynski et al. (2007a). 
The detection of a pulsar orbiting a black hole would also help to constrain 
many features of compact binary and pulsar evolution.
This has been explored in Pfahl et al. (2005) and more recently in KH09 
in terms of millisecond pulsar-black hole systems.
Fortunately the models, owing to a greater number of possible observational comparisons, 
are more robust concerning predictions for coalescing 
NS-NS systems (as described above) and from this point on NS-NS systems are the
focus of this work.

\section{Long and short gamma-ray burst Galactic kinematics}
\label{s:grb}

\begin{figure}
  \includegraphics[width=84mm]{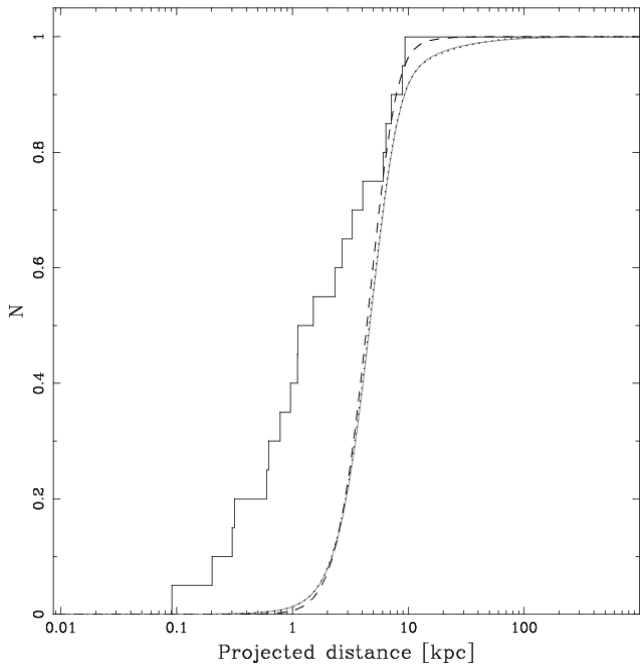}
  \caption{
    The projected distance from host galaxy for our different 
    double compact systems that coalesce (SGRB progenitors). 
    The projected distance is $\pi/4$ times the Model C$^{'''}$ 
    Galactic radial distance (as in Voss \& Tauris 2003).
    Full grey line gives NS-NSs, dashed line is BH-BHs 
    and the dotted line is the two NS and BH combinations.
    We note that to model SGRBs via double compact coalescence
    requires that the progenitor system must contain at 
    least one NS, for the formation of an accretion disk
    (see Voss \& Tauris 2003 and references therein).
    The LGRB observations of Bloom, Kulkarni \& Djorgovski (2002)
    are depicted by the jagged histogram.
    \label{f:fig19NS-NS}}
\end{figure}

\begin{figure}
  \includegraphics[width=84mm]{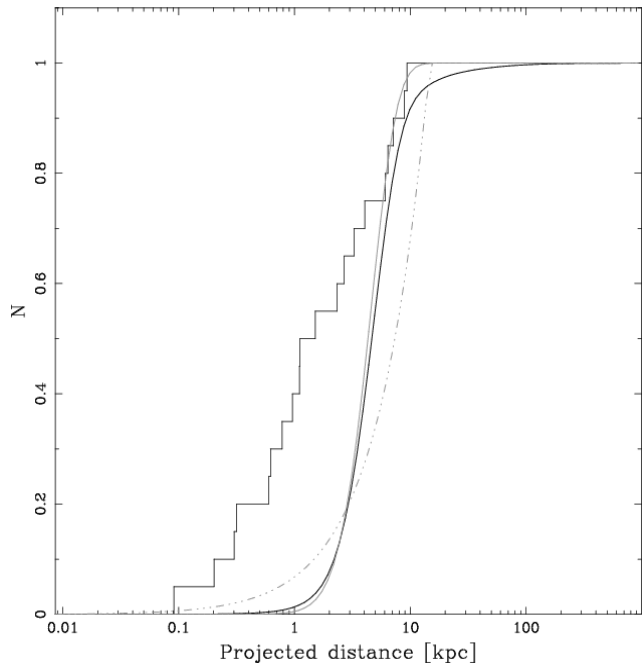}
  \caption{
    The projected distance from host galaxy for our different 
    GRB progenitors.
    The projected distance is $\pi/4$ times the Model C$^{'''}$ 
    GRB radial distances (as in Voss \& Tauris 2003).
    Full black line gives NS-NS-SGRBs, 
    full grey line is collapsar-LGRBs assuming the birth
    distribution of Yuslifov \& Kucuk (2004) and
    dashed-dotted grey line is the collapsar-LGRB
    distribution assuming a flat Galactic radial
    distribution between $0.01 < R < 20~$kpc.
    The observations of Bloom, Kulkarni \& Djorgovski (2002)
    are depicted by the jagged histogram.
    \label{f:fig21NS-NS}}
\end{figure}

We now examine the Galactic spatial distributions for 
our model GRBs, which require the following assumptions.
We assume the coalesced NS-NSs which result in a BH 
are short gamma-ray bursts while our collapsar model 
(see Section~\ref{s:bpbk}) is assumed
to form the long duration gamma-ray bursts.
Figure~\ref{f:fig19NS-NS} shows projected distances from the
host galaxy center for our Model C$^{'''}$ 
coalescing DCB populations.
The curves for all coalescing DCB system types are provided,
noting that BH-BH mergers can not form gamma-ray bursts.
We see that the different SGRB progenitor populations all 
follow very similar curves. 
The only slight difference is the BH-BH systems which receive 
reduced recoil velocities on average 
(see Figures~\ref{f:fig12NS-NS} and~\ref{f:fig15NS-NS})
causing them to coalesce slightly 
closer to their original radial birth location compared
to the other system types.
Importantly, even the NS-NS cumulative curve
follows the stellar radial birth cumulative 
curve (not shown) closely.
At first glance such a result is surprising: previously
it has been argued that because NSs receive large 
recoil velocities (Hobbs et al. 2005) SGRBs should
not typically appear in star forming regions 
(Bloom, Kulkarni \& Djorgovski 2002; Woosley \& Bloom 2006). 
This is in contrast to LGRBs whose progenitors are thought 
to be massive stars -- young stellar systems -- which should 
correlate well with star forming regions 
(Bloom, Kulkarni \& Djorgovski 2002). 
Yet owing to a population of close double compact
systems, and thus a large population of systems that coalesce
rather promptly (see Figure~\ref{f:fig182NS-NS} which is 
in line with those produced in Belczynski, 
Bulik \& Rudak 2002 and Voss \& Tauris 2003),
many of the model SGRBs merge at similar radial 
positions to their birth positions.
As noted in Section~\ref{s:coal} this feature of 
coalescing NS-NSs is also responsible for the smaller
scale height of coalescing NS-NSs compared to that of
BH-NS coalescence (see Table~\ref{t:table3}). 

As shown in KH09 the form of the host galaxy potential plays an 
important role in determining the resultant stellar kinematics. 
The Galactic potential of our model, taken from Paczy\'{n}ski (1990), 
is more dominant than that used in Voss \& Tauris (2003) but we 
do not see a significant difference in the SGRB distributions. 
We find that our model projected distances compare best with 
model f of Bloom, Sigurdsson \& Pols (1998) which assumes a 
similar circular velocity ($225\,$km s$^{-1}$ compared to our 
$220\,$km s$^{-1}$) and an identical SN velocity kick dispersion 
of $190~$km s$^{-1}$. 

In Figure~\ref{f:fig19NS-NS} we compare our SGRB models
to observations of projected LGRB distances from their
host galaxy centers according to Bloom, Kulkarni \& 
Djorgovski (2002).
We see that all 
Model C$^{'''}$ coalescing DCB populations fail to 
reproduce the LGRB observations. 
This is not exactly a surprise and agrees with previous 
studies 
(e.g. Bloom, Sigurdsson \& Pols 1999; 
Belczynski, Bulik \& Rudak 2002; Voss \& Tauris 2003). 
What we emphasise here is that the difference does not arise 
from SNe recoil velocities but instead could be telling us something 
about the assumed initial birth distribution of the stars and binaries. 
In our model the birth distribution is based on observations, in 
particular the distributions of OB stars and HII in the Galaxy as determined 
by Yusifov \& Kucuk (2004). 
Of course, at this stage, we are reliant upon 
making comparisons of two different 
populations, observed long gamma-ray bursts and 
model short gamma-ray bursts, and we note that 
statistics of observed SGRBs do not yet allow 
for meaningful comparisons (Savaglio, Glazebrook \& Le Borgne 2009). 
Another factor is that our models so far have been for 
the Milky Way whereas the LGRB observations are 
predominantly from dwarf galaxies 
(see Figure~\ref{f:fig22NS-NS} and associated text). 

To understand the GRB distributions further we next 
provide results of LGRB models (rather than
our previous SGRB models) to compare with the
observations of \textit{long}-GRBs (Bloom, Kulkarni 
\& Djorgovski 2002). 
The method of determining LGRB progenitors is outlined 
in Section~2 and the important point to note is that the 
average age of each system (at the time of the LGRB) is 
young, less than $\sim 10\,$Myr, with little dependence of 
the formation mechanism on uncertain features of binary evolution. 
So it is safe to assume that the LGRBs will trace the birth positions 
of their progenitors very closely. 
This is distinct from the SGRB models where the lifetimes of 
coalescing systems  depend upon a number of 
evolutionary features and assumptions (common-envelope evolution 
and SNe velocity kicks, for example) as 
discussed in detail in Sections~\ref{s:DCBNS-NS} and~\ref{s:coal} and
Belczynski, Bulik \& Rudak (2002).

\begin{figure}
  \includegraphics[width=84mm]{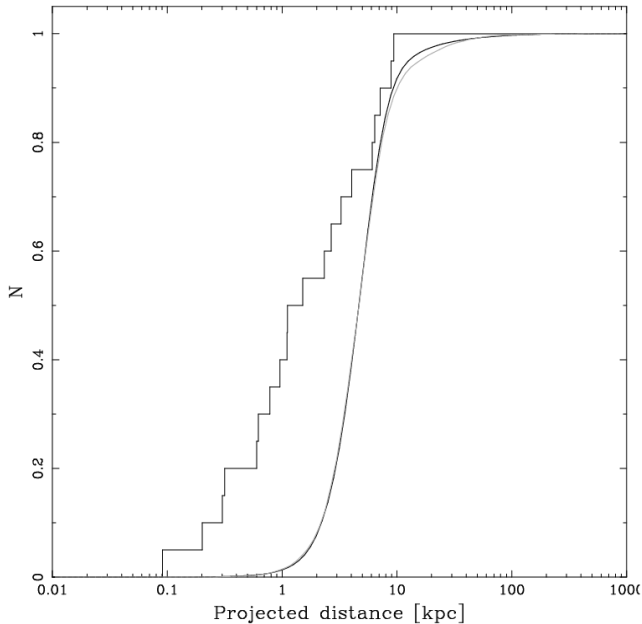}
  \caption{
    The projected distance from host galaxy for SGRBs 
    of different models.
    The projected distance is $\pi/4$ times the Model C$^{'''}$ 
    GRB radial distances (as in Voss \& Tauris 2003).
    The grey line is the SGRB  
    distribution assuming $\alpha_{\rm CE} = 1$ and the 
    full black line is Model C$^{'''}$ SGRBs.
    The observations of Bloom, Kulkarni \& Djorgovski (2002)
    are depicted by the jagged histogram.
    \label{f:alphaComparison}}
\end{figure}

The model LGRB projected distance distribution is 
compared in Figure~\ref{f:fig21NS-NS} to the LGRB 
observations and our previous NS-NS-SGRB distribution. 
This clearly shows that both our model
GRBs -- long and short -- cannot reproduce the observations
of GRB distances from their host galaxies. 
This is the case even when we allow an unrealistic radial 
birth description that is flat in Galactic radius 
(dashed-dotted line of Figure~\ref{f:fig21NS-NS}).
Assuming a flat radial birth distribution does increase 
the number of LGRBs that occur in the inner Galactic 
regions compared to our standard LGRB and SGRB models (though
not enough compared to observations), 
however, it also produces too many GRBs beyond 
$\sim 2-4~$kpc compared to the other models.
Hence, the collapsar model requires birth placement which
is incompatible with Population I stars in Milky Way like
galaxies.

\begin{figure}
  \includegraphics[width=84mm]{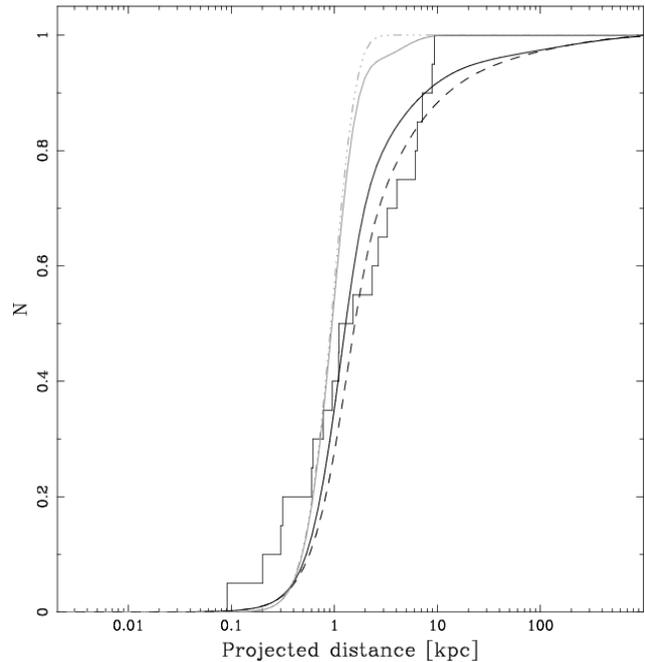}
  \caption{
    The projected distance from host dwarf galaxy for 
    our different GRB progenitor systems.
    The projected distance is $\pi/4$ times the model 
    radial distances (as in Voss \& Tauris 2003).
    The dashed-dotted grey line is the collapsar-LGRB 
    distribution assuming a scaled down model of our Galaxy
    ($\alpha_{\rm g} = 0.01$).
    Full black line and the full grey line 
    gives the NS-NS-SGRBs and collapsar-LGRBs 
    within a scaled spherical potential ($\alpha_{\rm g} = 0.1419$).
    The birth distribution for both dwarf galaxy models
    is that of Yuslifov \& Kucuk (2004) scaled by
    $\alpha_{\rm g} = 0.01$.
    The dashed line shows the NS-NS-SGRB distribution
    from a population of low metallicity stars ($Z = 0.0001$).
    The observations of Bloom, Kulkarni \& Djorgovski (2002)
    are depicted by the jagged histogram.
    \label{f:fig22NS-NS}}
\end{figure}

The majority of NS-NS systems merge rapidly when we assume
$\alpha_{\rm CE} = 1$, as shown in Figure~\ref{f:NS-NSmergertime}.
We compare this model SGRB projected distance distribution with LGRB observations and 
Model C$^{'''}$ in Figure~\ref{f:alphaComparison}.
The EC SN model is not shown as it does not differ greatly from Model C$^{'''}$
and has typically longer merger times -- which will not help here.
The only noticeable difference between Model C$^{'''}$ and our $\alpha_{\rm CE} = 1$
model is the distributions beyond about $10~{\rm kpc}$.
Here the greater relative number of systems with large merger times cause the distribution
to slope upward to unity slightly slower.
Unfortunately it is the other end of the distribution, at small distances, where
the models depart from observations.

This leads us to examine models of what is believed 
to be a more typical GRB host galaxy -- a dwarf 
galaxy. 
Dwarf galaxies contain less mass than the Galaxy, 
with masses of order $10^9~M_\odot$, and are 
typically modelled by simple potentials such 
as in Bloom, Sigurdsson \& Pols (1999).
In fact, 
Belczynski, Bulik \& Rudak found that 
the best match to GRB observations arose from a galaxy
mass of $0.01\times M_{\rm MW}$ -- the typical mass
of a LGRB host galaxy. 
Bloom, Sigurdsson \& Pols (1999) and Voss \& Tauris (2003)
also accounted for different galaxy masses, however, 
their initial stellar birth positions were not based on
observations of OB stars but instead followed
the exponential disk potentials -- something which can
be estimated from stellar kinematics 
(e.g. Kuijken \& Gilmore 1989) and light profiles 
(e.g. de Vaucouleurs 1948).
However, such observations measure stellar 
systems that differ from GRB progenitors.

To examine the effect such a small mass galaxy and 
gravitational potential has on our LGRB and SGRB 
populations we have produced two new galaxy models.
The first model is a scaled down version of our Galaxy 
where, in line with Bloom, Sigurdsson \& Pols (1999), 
Belczynski, Bulik \& Rudak (2002) and Voss \& Tauris (2003), 
we scale our previous model Galaxy mass, as described in
KH09, by the scale parameter $\alpha_{\rm g}$, giving us 
$M_{\rm galaxy} = \alpha_{\rm g} M_{\rm MW}$.
Assuming a constant density distribution between
models,
\begin{equation}
\alpha_{\rm g} M_{\rm MW} \propto R_{MW}^3,
\end{equation}
we then rescale the model scale lengths, scale heights and 
initial birth positions, also described in KH09, by 
$\alpha_{\rm G}^{1/3}$.
This first dwarf galaxy model assumes $\alpha_{\rm g} = 0.01$
(giving us a similar galaxy mass and size of the favoured 
galaxy model in the work of Belczynski, Bulik \& Rudak 2002),
from which the mass of the dwarf galaxy is
$M_{\rm galaxy} = 1.419\times 10^9~M_\odot$.
The second model takes 
a spherical Hernquist (1990) potential as used
in KH09 but scaled so the galaxy mass is the
same as in our first dwarf galaxy model, thus
$\alpha_{\rm g} = 0.1419$.
However, we keep $\alpha_{\rm g} = 0.01$ when scaling the
birth distribution in galactic $z$ and $R$. 
We then evolve our SGRB and LGRB progenitor systems in each of the 
dwarf galaxy models. 
The resultant projected distance curves are 
compared with the observations of 
Bloom, Kulkarni \& Djorgovski (2002) in 
Figure~\ref{f:fig22NS-NS} 
(note we do not plot the SGRB model for the first scenario). 
All populations now more closely
fit the observations than those in Figure~\ref{f:fig21NS-NS},
however, our LGRB models seem to over-estimate the number 
of GRB systems displaced between $1-10~$kpc from the host 
galaxy central region and under-estimate the number of 
GRB systems displaced less than $1~$kpc from the 
host galaxy.
It is interesting that the best fit to the observations
arises from our SGRB model, although this also 
under-estimates the number of GRB systems that occur 
in the inner galactic region.

The observed GRB host galaxies span a great 
range in red shift ($0 < z < 6.3$: Savaglio, 
Glazebrook \& Le Borgne 2009) 
and as such many GRB host galaxies are low 
metallicity galaxies (Fruchter et al. 2006) -- the average 
host galaxy metallicity being roughly one quarter 
solar (Savaglio, Glazebrook \& Le Borgne 2009).
The initial stellar metallicity has an important
effect on governing the progenitor mass range
for the formation of compact stars, in particular
we note that NSs may emerge from lower mass 
stars in low metallicity environments 
than could possibly occur for higher 
metallicity populations.
Therefore, compared to solar metallicity galaxies the 
resulting stellar and binary evolution in low metallicity
galaxies could possibly extend the lifetimes of pre- and 
post-NS-NS formation, allowing such systems more time to evolve
kinematically.
This motivated us to examine what effect a low
metallicity environment would have on the resultant
SGRB population.
To complete this we produced a SGRB model in our spherical
Hernquist potential and assumed a metallicity of $Z = 0.0001$, which is 
depicted within Figure~\ref{f:fig22NS-NS}.
Even with such a large difference in metallicity 
of the two SGRB models shown there does not appear to be 
a significant difference between their projected distance distributions.
The number of model systems found close to the host galaxy
(projected distance $< 1~$kpc) is still underestimated in this
model although the slope of the distribution does become less steep. 
Although this suggests that metallicity does not play a major role in 
kinematics of SGRBs (and it does not within the assumptions used here),
if mass loss rates are found to vary strongly with Z 
(Vink, de Koter \& Lamens, 2001; Belczynski et al. 2010) and the kick strength 
prescription is allowed to vary with mass of the exploding star 
(e.g. from fallback as assumed in Belczynski et al. 2008)
then there could indeed be a greater level of dependence on metallicity 
for SGRB kinematics.

Our models appear to confirm current observations 
that LGRBs are born in star forming regions
(Kelly, Kirshner \& Pahre, 2008; Dado \& Dar 2009).
We find that the final model GRB projected distance 
distributions are close to their respective birth
distributions, while, when compared to observations, 
we under-estimate the number of GRBs that occur in 
the inner galactic region.
Recently Fruchter et al. (2006) have suggested 
that there is a difference in LGRB and core collapse SN 
galactic environments.
This could indicate that there is a difference in
the birth distributions of the progenitors of these 
systems or that they arise from the same star formation 
distribution but from distinct progenitor mass ranges.
Within our models we use the same birth distribution 
for both our LGRB and SGRB progenitors, namely the 
distribution according to Yusifov \& Kucuk (2004) which 
should be realistic for SN progenitors and indeed was 
shown in KH09 to work well for pulsar models.
However, as observations of the Galaxy suggest
(Muno et al. 2006a; Muno et al. 2006b) there appears
to be a relatively large population of massive
stars within the inner $\sim 300~$pc.
This may be the case for many galaxies and suggests
that the Yusifov \& Kucuk (2004) radial birth 
distribution is inappropriate for modelling
the central galactic region.
Thus it is quite possible that massive stellar systems
may be born with a different galactic birth
distribution than standard stars.
In particular, the massive stellar birth distribution may 
be a combination of the the typical stellar distribution 
and a more centralised distribution.
This scenario would suit both our models and 
observations that the material surrounding LGRBs is
being highly irradiated by approximately $10-100$ times
more intensive UV radiation than that of the solar 
neighbourhood (Tumlinson et al. 2007) 
and that LGRBs occur in relatively low 
stellar density regions (Hammer et al. 2006; 
Tumlinson et al. 2007). 
It is also possible that 
this difference in the radial 
birth distribution may arise from open clusters 
with more massive stellar components falling quickly 
into the central galactic regions and becoming tidally 
disassociated (as in Muno et al. 2006b) or perhaps
these more massive stars are formed from a metal poor
distribution, allowing the birth of more massive stars
that can evolve in isolation and still cause a LGRB 
(Yoon, Langer \& Norman 2006).

The issue of whether or not LGRB and SGRB progenitors
follow the same radial birth distributions has important
consequences for the role that population synthesis studies
can play in understanding GRB observations.
If they do follow the same birth distribution then population
synthesis models can not reliably distinguish between 
LGRB and SGRB progenitors via the resultant GRB-galaxy off-sets. 
This is because both GRB populations trace star 
forming regions (although given time SGRBs can also 
occur in old stellar systems).
However, if the two GRB populations
have different galactic birth distributions, 
as raised in the above discussions, then further 
modelling of GRB populations can figure prominently in 
shedding light on the continuing GRB progenitor problem.

\section{Summary}

This work investigated the stellar, binary and 
Galactic kinematical evolutionary features of double
compact binary systems and possible short gamma-ray burst 
(coalescing NS-NSs) and long gamma-ray burst (tidally influenced 
collapsars) objects.
The main conclusions from this investigation are 
summarised here:
\begin{itemize}

\item NS-NS systems are typically more kinematically active and have 
the greatest scale height of all DCBs, eclipsing the BH-BH population
especially when BHs are assumed not to receive velocity kicks at birth.
Including electron capture supernovae into models alters this
somewhat.
When assuming electron capture supernovae occur and that
BHs receive kicks the BH-BH population has a comparable scale 
height to NS-NSs.

\item We find the double neutron-star formation rate 
ranges between $4~$Myr$^{-1}$ and $162~{\rm Myr}^{-1}$.
The lower value is set by assuming $\alpha_{\rm CE} = 1$,
the upper value by assuming EC SNe occurs. 
Our model without EC SNe and with $\alpha_{\rm CE} = 3$ 
gives a double neutron-star formation rate of 
$38~{\rm Myr}^{-1}$.

\item Our model double black-hole formation rate
ranges between $42~{\rm Myr}^{-1}$ and $820~{\rm Myr}^{-1}$.
The lower value arises when we assume BHs receive kicks,
the upper value arises when we do not assume BHs receive kicks.

%
\item If NS-NSs merge it is likely they do so within a time 
scale of a few million years and with a merger rate that ranges
between $3~$Myr$^{-1}$ and $154~{\rm Myr}^{-1}$ for our models.
The double neutron-star merger rates closely reflect their birth rates. 

\item BH-BH systems typically merge more slowly than NS-NS systems
and this difference is enhanced when BHs do not receive kicks.
The range of merger rates for double black holes is $25~{\rm Myr}^{-1}$
to $107~{\rm Myr}^{-1}$. 
The most limiting factor is the addition of SN kicks to BH formation, 
while assuming $\alpha_{\rm CE} = 1$ results in a merger
rate of $56~{\rm Myr}^{-1}$.
The maximum double BH merger rate arises when we assume they
do not feel kicks at birth.
Such a high merger rate suggest that at least one merger event 
should have been detected with LIGO.
The null detection of gravitational waves by LIGO gives credence
to the latest findings that BHs should receive kicks at birth.

\item Common envelope evolution is reconfirmed to be a very
important process for DCB formation.
In particular, merger time scales are sensitive to the number of CE
phases that occur within the intervening time between SN events.
Decreasing the efficiency of the CE process decreases the
typical merger time scale and final number of NS-NS systems.
The CE phase is also important in shaping the final 
NS-NS eccentricity-orbital period distribution.

\item We find good agreement of the shape of eccentricity-orbital 
period distribution when compared to observations but poor 
agreement with observations on the relative number of high 
eccentricy to low eccentricy systems.
Including electron capture SN evolution into population synthesis
models does not rectify the situation, as has been suggested previously.

%
\item We find poor agreement between the projected distance of 
long gamma-ray bursts from their host galaxy and the 
projected distances of NS-NS systems within a Milky Way model.

\item We find much better agreement between observations and
models when the model galaxy mass and size is scaled down by
a factor $\alpha_{\rm G} = 0.01$.
Owing to the short life times of our model systems (both short 
gamma-ray burst and long gamma-ray burst) 
the final projected distances depend heavily on 
the assumed radial birth distribution.


\end{itemize}

\section*{Acknowledgements}
PDK thanks Swinburne University of Technology for a
PhD scholarship.
The authors thank the referee for help with this paper.

\bsp

\label{lastpage}


\begin{thebibliography}{}
\bibitem[\protect\citeauthoryear{Abramovici et al.}{1992}]{Abr92}
    Abramovici, A., et al. 1992, Nature, 256, 325
\bibitem[\protect\citeauthoryear{Anderson et al.}{1990}]{And90}
    Anderson S.B., Gorham P.W., Kulkarni S.R., Prince T.A., Wolszczan A., Nature, 1990, 346, 42
\bibitem[\protect\citeauthoryear{Belczynski,Bulik \& Rudak}{2002}]{Bel02a}
    Belczynski K., Bulik T., Rudak B., 2002, ApJ, 571, 394
\bibitem[\protect\citeauthoryear{Belczynski, Kalogera \& Bulik}{2002}]{Bel02b}
    Belczynski K., Kalogera V., Bulik T., 2002, ApJ, 572, 407
\bibitem[\protect\citeauthoryear{Belczynski et al.}{2007a}]{Bel07a}
    Belczynski K., Taam R.E., Kalogera V., Rasio F.A., Bulik T., 2007a, ApJ, 662, 504
\bibitem[\protect\citeauthoryear{Belczynski et al.}{2007b}]{Bel07b}
    Belczynski K., Bulik T., Heger A., Fryer C., 2007b, ApJ, 664, 986
\bibitem[\protect\citeauthoryear{Belczynski et al.}{2008}]{Bel08}
    Belczynski K., Kalogera V., Rasio, F.A., Taam R.E., Zezas A., Bulik T., Maccarone T.J., Ivanova N., 2008, ApJS, 174, 223
\bibitem[\protect\citeauthoryear{Belczynski et al.}{2010}]{Bel10}
    Belczynski K., Tomasz B., Fryer C.L., Ruiter A., Valsecchi F., Vink J.S., Hurley J., 2010, accepted ApJ, astro-ph/0904.2784B
\bibitem[\protect\citeauthoryear{Bloom, Sigurdsson \& Pols}{1999}]{Blo99}
    Bloom J.S., Sigurdsson S., Pols O.R., MNRAS, 305, 763
\bibitem[\protect\citeauthoryear{Bloom, Kulkarni \& Djorgovski}{2002}]{Blo02}
    Bloom J.S., Kulkarni S.R., Djorgovski S.G., 2002, ApJ, 123, 1111
\bibitem[\protect\citeauthoryear{Bodenheimer \& Taam}{1984}]{Bod84}
    Bodenheimer P., Taam R.E., 1984, ApJ, 280, 771
\bibitem[\protect\citeauthoryear{Bogomazov, Lipunov \& Tutukov}{2008}]{Bog08}
    Bogomazov A.I., Lipunov V.M., Tutukov A.V., 2008, Astronomy Reports, 53, 463
\bibitem[\protect\citeauthoryear{Brown}{1995}]{Bro95}
    Brown G.E., 1995, ApJ, 440, 270
\bibitem[\protect\citeauthoryear{Burgay et al.}{2003}]{Bur03}
    Burgay M., et al., 2003, Nature, 462, 531
\bibitem[\protect\citeauthoryear{Cappellaro, Evans \& Turatto}{1999}]{Cap99}
    Cappellaro E., Evans R., Turatto M., 1999, A\&A, 351, 459
\bibitem[\protect\citeauthoryear{Champion et al.}{2004}]{Cha04}
    Champion D.J., Lorimer D.R., McLaughlin M.A., Cordes J.M., Taylor J.H., 2004, MNRAS, 350, L61
\bibitem[\protect\citeauthoryear{Champion et al.}{2008}]{Cha08}
    Champion D.J., et al., 2008, Science, 320, 1309
\bibitem[\protect\citeauthoryear{Chaurasia \& Bailes}{2005}]{Cha05}
    Chaurasia H.K., Bailes M., 2005, ApJ, 632, 1054 
\bibitem[\protect\citeauthoryear{Clark \& Eardley}{1977}]{Cla77}
    Clark J.P.A., Eardley D.M., 1977, ApJ, 251, 311
\bibitem[\protect\citeauthoryear{Dado \& Dar}{2009}]{Dad09}
    Dado S., Dar A., 2009, ApJ, 693, 311
\bibitem[\protect\citeauthoryear{Detmers et al.}{2008}]{Det08}
    Detmers R.G., Langer N., Podsiadlowski Ph, Izzard R.G., 2008, A\&A, 484, 831
\bibitem[\protect\citeauthoryear{Dewi \& Tauris}{2001}]{Dew01}
    Dewi J. D. M., Tauris T. M., 2001, in Podsiadlowski P., Rappaport S., King
A. R., D’Antona F., Burderi L., eds, ASP Conf. Ser. Vol. 229, Evolution
of Binary and Multiple Star Systems. Astron. Soc. Pac., San Francisco,
p. 255
\bibitem[\protect\citeauthoryear{Dewi, Podsiadlowski \& Sena}{2006}]{Dew06}
    Dewi J. D. M., Podsiadlowski Ph., Sena A., 2006, MNRAS, 368, 1742
\bibitem[\protect\citeauthoryear{de Vaucouleurs}{1948}]{deV48}
    de Vaucouleurs G., 1984, AnAp, 11, 247
\bibitem[\protect\citeauthoryear{Fruchter et al.}{2006}]{Fru06}
    Fruchter A.S., et al., 2006, Nature, 441, 463
\bibitem[\protect\citeauthoryear{Faulkner et al.}{2005}]{Fau05}
    Faulker A.J., et al. 2005, ApJ, 618, 119
\bibitem[\protect\citeauthoryear{Galloway}{2008}]{Gal08a}
    Galloway D.K., 2008, in Bassa C.G., Wang Z., Cumming A., Kaspi V.M., eds., AIP Conf. Ser. Vol. 938, 40 Years of Pulsars -- Millisecond Pulsars, Magnetars, and More, Americal Inst. of Physics, New York, p. 510
\bibitem[\protect\citeauthoryear{Galloway et al.}{2008}]{Gal08b}
    Galloway D.K., Muno M.P., Hartman J.M., Psaltis D., Chakrabary D., 2008, ApJS, 179, 360
\bibitem[\protect\citeauthoryear{Hammer}{1892}]{Ham92}
    Hammer E., 1892, Petermanns Mitt., 38, 85
\bibitem[\protect\citeauthoryear{Hammer et al.}{2006}]{Ham06}
    Hammer F., Flores H., Schaere D., Dessauges-Zavadsky M., Le Floc'h E., Puech M., 2006, A\&A, 454, 103 
\bibitem[\protect\citeauthoryear{Han, Podsiadlowski \& Eggleton}{1995}]{Han95}
    Han Z., Podsiadlowski P., Eggleton P.P., 1995, MNRAS, 272, 800
\bibitem[\protect\citeauthoryear{Hernquist}{1990}]{Her90}
    Hernquist L., 1990, ApJ, 356, 359
\bibitem[\protect\citeauthoryear{Hobbs et al.}{2005}]{Hob05}
    Hobbs G., Lorimer D.R., Lyne A.G., Kramer M., 2005, MNRAS, 315, 543
\bibitem[\protect\citeauthoryear{Hulse \& Taylor}{1975}]{Hul75}
    Hulse R.A., Taylor J.H., 1975, ApJ, 195, 51L
\bibitem[\protect\citeauthoryear{Hurley, Tout \& Pols}{2002}]{Hur02}
    Hurley J.R., Tout C.A., Pols O.R., 2002, MNRAS, 329, 897
\bibitem[\protect\citeauthoryear{Iben \& Livio}{1993}]{Ibe93}
    Iben I.Jr., Livio M., 1993, PASP, 105, 1373
\bibitem[\protect\citeauthoryear{Ihm, Kalogera \& Belczynski}{2006}]{Ihm06}
    Ihm C.M., Kalogera V., Belczynski K., 2006, ApJ, 652, 540
\bibitem[\protect\citeauthoryear{Ivanova \& Taam}{2003}]{Iva03}
    Ivanova N., Taam R.E., 2003, ApJ, 599, 516
\bibitem[\protect\citeauthoryear{Ivanova et al.}{2008}]{Iva08}
    Ivanova N., Heinke C.O., Rasio F.A., Belczynski K., Fregeau J.M., 2008, MNRAS, 386, 553
\bibitem[\protect\citeauthoryear{Jonker \& Nelemans}{2004}]{Jon04}
    Jonker P.G., Nelemans G., 2004, MNRAS, 354, 355
\bibitem[\protect\citeauthoryear{Kalogera}{1996}]{Kal96}
    Kalogera V., 1996, ApJ, 471, 352
\bibitem[\protect\citeauthoryear{Kalogera}{1998}]{Kal98}
    Kalogera V., 1998, ApJ, 493, 368
\bibitem[\protect\citeauthoryear{Kalogera et al.}{2007}]{Kal07}
    Kalogera V., Belczynski K., Kim C., O'Shaughnessy R., Willems B., 2007, PhR., 442, 75
\bibitem[\protect\citeauthoryear{Kiel \& Hurley}{2006}]{Kie06}
    Kiel P.D., Hurley J.R., 2006, MNRAS, 369, 1152
\bibitem[\protect\citeauthoryear{Kiel, Hurley, Murray, Hayasaki}{2007}]{Kie07}
    Kiel P.D., Hurley J.R., Murray J.R., Hayasaki K., 2007, in Stefl S., Owocki S.P., Okazaki A.T., eds., ASP Conf. Ser. Vol. 361, Active OB-Stars: Laboratories For Stellar and Circumstellar Physics. ASP, San Francisco, p. 448
\bibitem[\protect\citeauthoryear{Kiel, Hurley, Bailes \& Murray}{2008}]{Kie08}
    Kiel P.D., Hurley J.R., Bailes M., Murray J.R., 2008, MNRAS, 388, 393
\bibitem[\protect\citeauthoryear{Kiel \& Hurley}{2009}]{Kie09b}
    Kiel P.D., Hurley J.R., 2009, MNRAS, 402, 1437 (KH09)
\bibitem[\protect\citeauthoryear{Kim, Kalogera \& Lorimer}{2006}]{Kim06}
    Kim C., Kalogera V., Lorimer D.R., 2006, in Kapers L., van der Klis M., Wijers R., New Ast. Rev., A Life with Stars. Elsevier, Amsterdam, in press, astroph/0608280v1
\bibitem[\protect\citeauthoryear{Kelly, Kirshner \& Pahre}{2008}]{Kel08}
    Kelly P.L., Kirshner R.P., Pahre M., 2008, ApJ, 687, 1201
\bibitem[\protect\citeauthoryear{Kroupa, Tout \& Gilmore}{1993}]{Kro93}
    Kroupa P., Tout C.A., Gilmore G., 1993, MNRAS, 262, 545
\bibitem[\protect\citeauthoryear{Kuijken \& Gilmore}{1989}]{Kui89}
    Kuijken K., Gilmore G., 1989, MNRAS, 239, 571
\bibitem[\protect\citeauthoryear{Landau \& Lifshitz}{1951}]{Lan51}
    Landau L.D., Lifshitz E.M., 1951, The classical Theory of Fields, 1st English edn. Pergamon Press, Oxford
\bibitem[\protect\citeauthoryear{Lazzarini}{2007}]{Laz07}
    Lazzarini A., 2007, Update from LIGO laboratory, LIGO Document G070649-00-M
\bibitem[\protect\citeauthoryear{Liu, van Paradijs \& van den Heuvel}{2007}]{Liu07}
    Liu Q.Z., van Paradijs J., van den Heuvel E.P.J., 2007, A\&A, 469, 807
\bibitem[\protect\citeauthoryear{Lorimer et al.}{2006a}]{Lor06a}
    Lorimer D.R., et al., 2006a, MNRAS, 372, 777
\bibitem[\protect\citeauthoryear{Lorimer et al.}{2006b}]{Lor06b}
    Lorimer D.R., et al., 2006b, ApJ, 640, 428
\bibitem[\protect\citeauthoryear{Lyne et al.}{2004}]{Lyn04}
    Lyne A.G., et al., 2004, Science, 303, 1153
\bibitem[\protect\citeauthoryear{Manchester et al.}{2005}]{Man05}
    Manchester R.N., Hobbs G.B., Teoh A., Hobbs M., 2005, ApJ, 129, 1993
\bibitem[\protect\citeauthoryear{MacFadyen \& Woosley}{1999}]{Mac99}
    MacFadyen A.I., Woosley S.E., 1999, ApJ, 524, 262
\bibitem[\protect\citeauthoryear{Miyaji et al.}{1980}]{Miy80}
    Miyaji S., Nomoto K., Yokoi K,. Sugimoto D., 1980, PASJ, 32, 303
\bibitem[\protect\citeauthoryear{Muno et al.}{2006a}]{Muno06a}
    Muno M.P., Bauer F.E., Bandyopadhayay R.M., Wang Q.D., 2006a, ApJSS, 165, 173
\bibitem[\protect\citeauthoryear{Muno et al.}{2006b}]{Muno06b}
    Muno M.P., Bower G.C., Burgasser A.J., Baganoff F.K., Morris M.R., Brandt W.N., 2006b, ApJ, 638, 183
\bibitem[\protect\citeauthoryear{Nelemans, \& Tout}{2005}]{Nel05}
    Nelemans G., Tout C.A., 2005, MNRAS, 356, 753
\bibitem[\protect\citeauthoryear{Nice, Sayer \& Taylor}{1996}]{Nic96}
    Nice D.J., Sayer R.W., Taylor J.H., 1996, ApJ, 466, 87L
\bibitem[\protect\citeauthoryear{O'Shaughnessy et al.}{2005a}]{OSa05}
    O'Shaughnessy R., Kim C., Kalogera V., Belczynski K., 2005, ApJ., 672, 479
\bibitem[\protect\citeauthoryear{O'Shaughnessy et al.}{2005b}]{OSb05}
    O'Shaughnessy R., Kim C., Fragos T., Kalogera V., Belczynski K., 2005, ApJ., 633, 1076
\bibitem[\protect\citeauthoryear{Paczy\'{n}ski}{1967}]{Pac67}
    Paczy\'{n}ski B., 1967, AcA, 17, 287
\bibitem[\protect\citeauthoryear{Paczy\'{n}ski}{1976}]{Pac76}
    Paczy\'{n}ski B., 1976, in Eggleton P., Mitton S., Whelan J., eds, Proc IAU Symp. 73, Structure and Evolution of Close Binary Systems. Reidel, Dordrecht, p. 75
\bibitem[\protect\citeauthoryear{Paczynski}{1986}]{Pac86}
    Paczy\'{n}ski B., 1986, ApJ, 308, 43
\bibitem[\protect\citeauthoryear{Paczynski}{1990}]{Pac90}
    Paczy\'{n}ski B., 1990, A\&A, 348, 485
\bibitem[\protect\citeauthoryear{Petrovic et al.}{2005}]{Pet05}
    Petrovic J., Langer N., Yoon S.-C., Heger A., 2005, A\&A, 435, 347
\bibitem[\protect\citeauthoryear{Pfahl, Rappaport \& Podsaidlowski}{2005}]{Pfa05}
    Pfahl E., Podsaidlowski P., Rappaport S., 2005, ApJ, 628, 343
\bibitem[\protect\citeauthoryear{Podsiadlowski, Rappaport \& Han}{2003}]{Pod03}
    Podsiadlowski Ph., Rappaport S., Han Z., 2003, MNRAS, 341, 385
\bibitem[\protect\citeauthoryear{Portegies Zwart \& Verbunt}{1996}]{Por96}
    Portegies Zwart S.F., Verbunt F., 1996, A\&A, 309, 179
\bibitem[\protect\citeauthoryear{Portegies Zwart \& Yungelson}{1998}]{Por98}
    Portegies Zwart S.F., Yungelson L.R., 1998, A\&A, 332, 173
\bibitem[\protect\citeauthoryear{Ricker \& Taam}{2008}]{Ric08}
    Ricker P.M., Taam R.E., 2008, ApJ, 672, 41
\bibitem[\protect\citeauthoryear{Sadowski et al.}{2008}]{Sad08}
    Sadowski A., Belczynski K., Bulik T., Ivanova N., Rasio F.A., O'Shaughnessy R., 2008, ApJ., 676, 1162
\bibitem[\protect\citeauthoryear{Savaglio, Glazebrook \& Le Borgne}{2009}]{Sav09}
    Savaglio S., Glazebrook K., \& Le Borgne D., 2009, ApJ, 691, 101
\bibitem[\protect\citeauthoryear{Stairs}{2004}]{Sta04}
    Stairs I.H., 2004, Science, 304, 547
\bibitem[\protect\citeauthoryear{Stairs}{2008}]{Sta08}
    Stairs I.H., 2008, in Bassa C.G., Wang Z., Cumming A., Kaspi V.M., eds., AIP Conf. Ser. Vol. 938, 40 Years of Pulsars -- Millisecond Pulsars, Magnetars, and More, Americal Inst. of Physics, New York, p. 424
\bibitem[\protect\citeauthoryear{Steers}{1970}]{Ste70}
    Steers J.A., 1970, An Introduction to the Study of Map Projections, University of London Press, London
\bibitem[\protect\citeauthoryear{Taam \& Sandquist}{2000}]{Taa00}
    Taam R.E., Sandquist E.L., 2000, ARA\&A, 38, 113
\bibitem[\protect\citeauthoryear{Tauris \& Manchester}{1998}]{Tau98}
    Tauris T.M., Manchester R.N., 1998, MNRAS, 298, 625
\bibitem[\protect\citeauthoryear{Tauris \& Dewi}{2001}]{Tau01}
    Tauris T.M., Dewi, J.D.M., 2001, A\&A, 369, 170
\bibitem[\protect\citeauthoryear{Taylor, Fowler \& McCullock}{1979}]{Tay79}
    Taylor J.H., Fowler L.A., McCulloch P.M., 1979, Nature, 277, 437
\bibitem[\protect\citeauthoryear{Taylor \& Manchester}{1977}]{Tay77}
    Taylor J.H., Manchester R.N., 1977, ApJ, 215, 885
\bibitem[\protect\citeauthoryear{Tumlinson et al.}{2007}]{Tum07}
    Tumlinson J, Prochaska J.X., Chen H-W., Dessauges-Zavadsky M., Bloom J.S., 2007, ApJ, 668, 667
\bibitem[\protect\citeauthoryear{Tutukov \& Yungelson}{1996}]{Tut96}
    Tutukov A., Yungelson C., 1996, MNRAS, 280, 1035
\bibitem[\protect\citeauthoryear{van den Heuvel}{2007}]{Van07}
    van den Heuvel E.P.J., 2007, in Antonelli A.L., et al. eds, AIP Conf. Proc. Vol. 942, The Multicolored Landscaep of Compact Objects and their Explosive Origins. Am. Inst. Phys., New York, p. 598
\bibitem[\protect\citeauthoryear{Vink, de Koter \& Lamens}{2001}]{Vin01}
    Vink J.S., de Koter A., Lamens H., 2001, A\&A, 369, 574
\bibitem[\protect\citeauthoryear{Voss \& Tauris}{2003}]{Vos03}
    Voss R., Tauris T.M., 2003, MNRAS, 342, 1169
\bibitem[\protect\citeauthoryear{Webbink}{1984}]{Web84}
    Webbink R.F., 1984, ApJ, 277, 355
\bibitem[\protect\citeauthoryear{Willems et al.}{2005}]{Wil05}
    Willems B., Henninger M., Levin T., Ivanova N., Kalogera V., McGhee K., Timmes F.X., Fryer C.L., 2005, ApJ, 625, 324
\bibitem[\protect\citeauthoryear{Woosley}{1993}]{Woo93}
    Woosley S.E., 1993, ApJ, 405, 273
\bibitem[\protect\citeauthoryear{Woosley \& Bloom}{2006}]{Woo06}
    Woosley S.E., Bloom J.S., 2006, Annu. Rev. Astron. Astrophys., 44, 507
\bibitem[\protect\citeauthoryear{Yoon, Langer \& Norman}{2006}]{Yoo06}
    Yoon S.-C., Langer N., Norman C., 2006, A\&A, 460, 199
\bibitem[\protect\citeauthoryear{Yusifov \& Kucuk}{2004}]{Yus04}
    Yusifov I., Kucuk I., 2004, A\&A, 422, 545

\end{thebibliography}
\end{document}